\documentclass[]{spie}  

\usepackage{amsmath,amsfonts,amssymb}
\usepackage{xcolor}
\usepackage{marginnote}

\usepackage{graphicx}
\usepackage[acronym]{glossaries}
\usepackage{array}
\usepackage{subcaption}
\usepackage{longtable,booktabs,array}
\usepackage{makecell}  
\usepackage{acronym}
\newacro{ccds}[CCDs]{charged coupled devices}
\newacro{lsstcam}[LSSTCam]{Legacy Survey of Space and Time Camera}
\newacro{cmb}[CMB]{Cosmic Microwave Background}
\newacro{lsst}[LSST]{Legacy Survey of Space and Time}
\newacro{ptc}[PTC]{photon-transfer curve}
\newacro{RTMs}[RTMs]{raft tower modules}
\newacro{RTM}[RTM]{raft tower module}
\newacro{e2v}[e2v]{Teledyne e2v}
\newacro{itl}[ITL]{Imaging Technology Laboratory}
\newacro{BNL}[BNL]{Brookhaven National Laboratory}
\newacro{UCD}[UC-Davis]{University of California, Davis}
\newacro{SLAC}[SLAC]{SLAC National Accelerator Laboratory}
\newacro{EOT}[EO testing]{electro-optical testing}
\newacro{TMA}[TMA]{Telescope Mount Assembly}
\newacro{S3DF}[S3DF]{\ac{SLAC} Shared Scientific Data Facility}
\newacro{CCOB}[CCOB]{Camera Calibration Optical Bench}
\newacro{TEA}[TEA]{top-end assembly}
\newacro{REB}[REB]{Readout Electronics Board}
\newacro{adc}[ADC]{analog to digital converter}
\newacro{PTC}[PTC]{Photon transfer curve}
\newacro{CTI}[CTI]{charge transfer inefficiency}
\newacro{eo_pipe}[\texttt{eo\_pipe}]{electro-optical data products pipeline}
\newacro{cp_pipe}[\texttt{cp\_pipe}]{calibration-products production pipeline}
\newacro{lsstcomcam}[LSSTComCam]{the LSST Commissioning Camera}
\newacro{RG}[RG]{Reset Gate}
\newacro{flats}[flats]{flat illumination images}

\usepackage[colorlinks=true, allcolors=blue]{hyperref}

\title{The On-Sky Performance of the LSST Camera CCD Array}

\author[a]{Sean Patrick MacBride}
\author[b]{Aaron Roodman}
\author[b]{Stuart Marshall}
\author[c]{Yousuke Utsumi}
\author[d]{Kevin Fanning}
\author[e]{John Banovetz}
\author[b]{Theo Schutt}
\author[b]{Alexander Broughton}
\author[d]{Shuang Liang}
\author[b]{Andrew P. Rasmussen}
\author[f]{Pierre Antilogus}
\author[d]{John Gregg Thayer}
\author[d]{Homer Neal}
\author[d]{Anthony S. Johnson}
\author[f]{Pierre Astier}
\author[d]{Seth W. Digel}
\author[d,b]{Andrew Bradshaw}
\author[g]{Johan Bregeon}
\author[b]{James Chiang}
\author[g]{C\'{e}line Combet}
\author[f]{Guillaume Dargaud}
\author[h]{Johnny H. Esteves}
\author[a]{Marcelle Soares-Santos}
\author[i]{Thibault Guillemin}
\author[f]{Claire Juramy-Gilles}
\author[j]{Craig S. Lage}
\author[g]{Myriam Migliore}
\author[j]{Daniel Polin}
\author[j]{J. Anthony Tyson}
\author[k]{Renee Nichols}
\author[k]{Steven M. Ritz}
\author[b]{Eli S. Rykoff}
\author[k]{Adrian Shestakov}
\author[j]{Adam Snyder}
\author[d]{Max Turri}
\author[k]{Duncan Wood}
\author[d]{Travis Lange}
\author[d]{Martin Nordby}
\author[d]{Hannah Mary Margaret Pollek}
\author[d]{Shawn Osier}
\author[d]{Boyd Bowdish}
\author[d]{Diane Hascall}
\author[d]{Margaux Lopez}
\author[d]{Scott P. Newbry}
\author[d]{Juan Carlos Lazarte}
\author[d]{Vincent Lee}
\author[d]{Michael Silva}
\author[d]{Dave Kiehl}
\author[d]{Andrew Hau}
\author[d]{Tom Nieland}
\author[d]{Nico Linton}
\author[d]{David Shelley}
\author[d]{Yongqiang (Brian) Qiu}
\author[d]{Mark Freytag}
\author[d]{Stephen Cisneros}
\author[d]{Chris Mendez}
\author[d]{Stephen A. Tether}
\author[k]{Alan M. Eisner}
\author[d]{Dmitry Onoprienko}
\author[d]{Owen Saxton}
\author[d]{Kevin A. Reil}
\author[l]{Vincent J. Riot}
\author[d]{Justin Wolfe}
\author[l]{Scott E. Winters}
\author[l]{Brian J. Bauman}
\author[e]{William Wahl}
\author[e]{Paul O'Connor}
\author[m]{Fran\c{c}oise Virieux}
\author[m]{Alexandre Boucaud}
\author[m]{Camille Parisel}
\author[n]{\'{E}ric Aubourg}
\author[m]{Eric Lagorio}
\author[o]{Pierre Karst}
\author[o]{Aur\'{e}lien Marini}
\author[f]{Didier Laporte}
\author[g]{Francis Vezzu}
\author[f]{Guillaume Daubard}
\author[o]{Patrick Breugnon}
\author[d]{Tim W. Bond}
\author[d,b]{Eric Charles}
\author[d]{Richard Dubois}
\author[p]{Steven M. Kahn}
\author[d,b]{Andr\'es A. Plazas Malag\'on}
\author[b]{Rafe H. Schindler}
\author[b]{Joanne Bogart}
\author[q]{Hye Yun Park}
\author[g]{Aur\'{e}lien Barrau}
\author[r]{Christopher Z. Waters}
\author[s]{Merlin Fisher-Levine}
\author[g]{Antoine Bernard}
\author[g]{Mile Kusulja}

\affil[a]{Physik-Institut, University of Zurich, Winterthurerstrasse 190, 8057 Zurich, Switzerland}
\affil[b]{Kavli Institute for Particle Astrophysics and Cosmology, SLAC National Accelerator Laboratory, 2575 Sand Hill Rd., Menlo Park, CA 94025, USA}
\affil[c]{National Astronomical Observatory of Japan, Chile Observatory, Los Abedules 3085, Vitacura, Santiago, Chile}
\affil[d]{SLAC National Accelerator Laboratory, 2575 Sand Hill Rd., Menlo Park, CA 94025, USA}
\affil[e]{Brookhaven National Laboratory, Upton, NY 11973, USA}
\affil[f]{Sorbonne Universit\'{e}, Universit\'{e} Paris Cit\'{e}, CNRS/IN2P3, LPNHE, 4 place Jussieu, F-75005 Paris, France}
\affil[g]{Universit\'{e} Grenoble Alpes, CNRS/IN2P3, LPSC, 53 avenue des Martyrs, F-38026 Grenoble, France}
\affil[h]{Department of Physics, Harvard University, 17 Oxford St., Cambridge MA 02138, USA}
\affil[i]{Universit\'{e} Savoie Mont-Blanc, CNRS/IN2P3, LAPP, 9 Chemin de Bellevue, F-74940 Annecy-le-Vieux, France}
\affil[j]{Physics Department, University of California, One Shields Avenue, Davis, CA 95616, USA}
\affil[k]{Santa Cruz Institute for Particle Physics and Physics Department, University of California--Santa Cruz, 1156 High St., Santa Cruz, CA 95064, USA}
\affil[l]{Lawrence Livermore National Laboratory, 7000 East Avenue, Livermore, CA 94550, USA}
\affil[m]{Universit\'{e} Paris Cit\'{e}, CNRS/IN2P3, APC, 4 rue Elsa Morante, F-75013 Paris, France}
\affil[n]{Universit\'{e} Paris Cit\'{e}, CNRS/IN2P3, CEA, APC, 4 rue Elsa Morante, F-75013 Paris, France}
\affil[o]{Aix Marseille Universit\'{e}, CNRS/IN2P3, CPPM, 163 avenue de Luminy, F-13288 Marseille, France}
\affil[p]{Physics Department,  University of California, 366 Physics North, MC 7300 Berkeley, CA 94720, USA}
\affil[q]{Department of Physics, Duke University, Durham, NC 27708, USA}
\affil[r]{Department of Astrophysical Sciences, Princeton University, Princeton, NJ 08544, USA}
\affil[s]{D4D CONSULTING LTD., Suite 1 Second Floor, Everdene House, Deansleigh Road, Bournemouth, UK BH7 7DU}

\authorinfo{*Author information: E-mail \href{mailto:sean.macbride@physik.uzh.ch}{sean.macbride@physik.uzh.ch}}

\begin{document} 
\maketitle

\begin{abstract}
The focal plane of the LSST Camera contains 189 individual science CCDs, arranged into 21 raft tower modules, along with 4 wavefront and 8 guider CCDs located in 4 additional corner RTMs. Altogether, the LSST Camera CCDs compose the largest focal plane ever constructed. The LSST Camera is the primary instrument of Rubin Observatory, which will begin the Legacy Survey of Space and Time in 2026. In this paper, we describe the on-sky performance of the LSST Camera CCDs, from receipt at NSF/DOE Vera C. Rubin Observatory in May 2024 to on-sky observations during the first year of operations. We discuss the process to establish functionality of several CCDs which were affected by an electrical short and faulty analog-digital converter, optimizations of readout timing in response to changes in the survey strategy, and implementation of enhanced focal plane safety measures through an active clearing mechanism on the CCDs. Finally, we discuss sensor features observed on-sky, and global performance during the first year of operations. The operations to date of the LSST Camera CCDs have demonstrated the capability of performing a wide, fast, and deep optical imaging survey of the entire southern sky at the Rubin Observatory.
\end{abstract}


\section{INTRODUCTION}
\label{sec:intro}\label{sec:introduction}
\begin{figure}[]
    \centering
     \includegraphics[width=\textwidth]{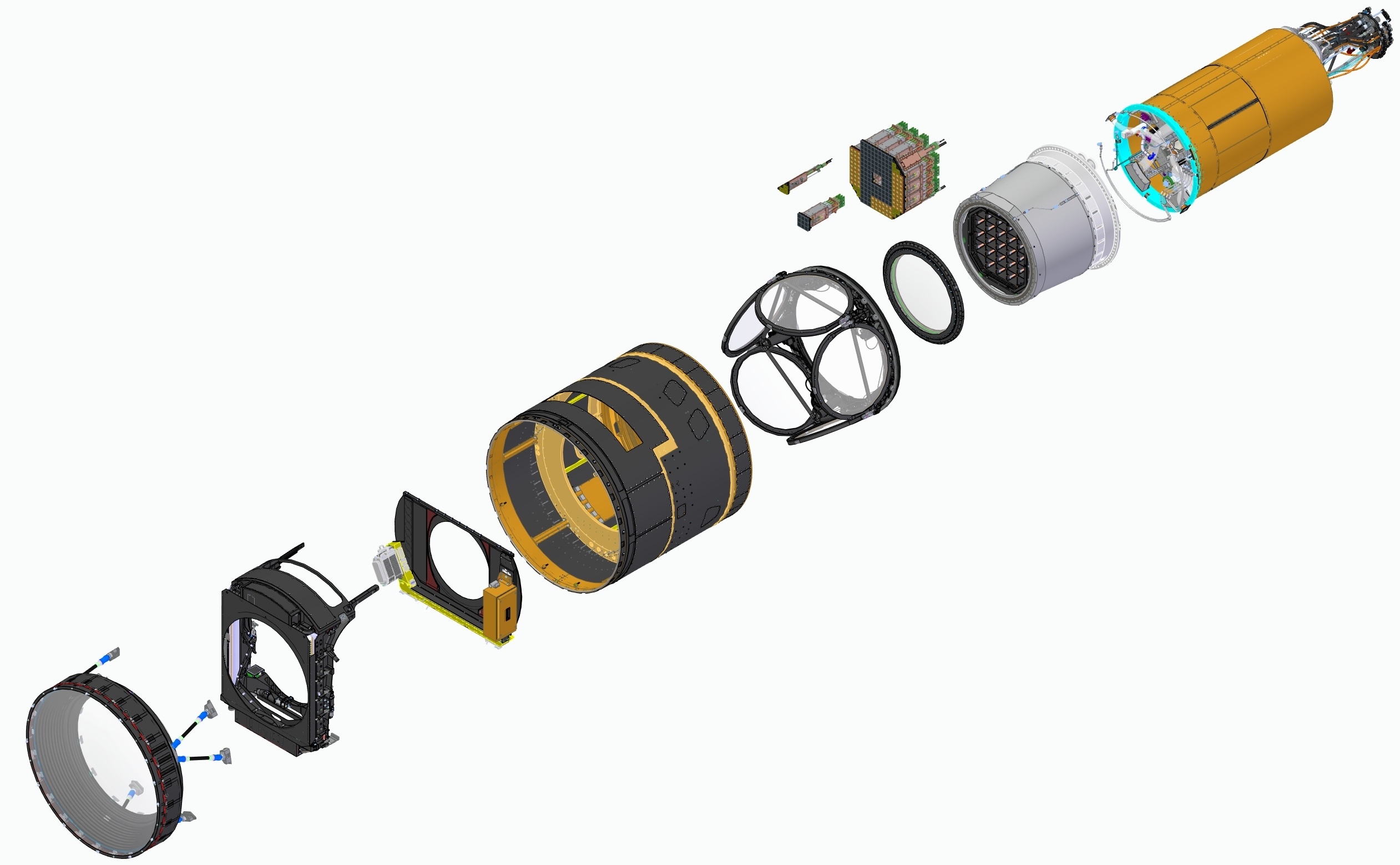}
     \caption{An exploded view of \ac{lsstcam}. Along the central axis, from left to right: the primary and secondary lenses of the optical corrector, the autochanger of the filter exchange system, the shutter, the camera body, the carousel of the filter exchange system, the cryostat window as the tertiary lens of the optical corrector, the cryostat body which the tertiary lens mounts to, and the utility trunk. The focal plane assembly is shown adjacent to the cryostat, with a detailed view in Figure \ref{fig:composite_layout}. For a detailed overview of the individual subsystems of \ac{lsstcam}, see Lange \cite{2024SPIE13096E..1OL}.}
     \label{fig:FullCamExplode2}
\end{figure}
The Vera C. Rubin Observatory represents a generational leap in wide-field optical astronomy, designed to execute the \ac{lsst}, a decade-long imaging program starting in 2026 that will repeatedly map the southern sky with unprecedented depth, cadence, and photometric stability.\cite{2019ApJ...873..111I} The \ac{lsst} will transform cosmology, time-domain astrophysics, solar system science, and our understanding of the Milky Way by producing a multi-band imaging dataset covering $\sim 18,000\deg^2$. Achieving these science goals depends on the on-sky performance of the primary instrument of Rubin Observatory, the \ac{lsstcam}.\cite{2024SPIE13096E..1SR}

\ac{lsstcam} is the largest digital camera ever built, with a 3.2 gigapixel focal plane. The focal plane is contained within the cryostat, one of the many subsystems of \ac{lsstcam}. The primary subsystems of \ac{lsstcam} are the focal plane assembly, cryostat, optical corrector, cryostat, filter exchange system, utility trunk and auxiliary electronics, and camera shutter, all shown in Figure \ref{fig:FullCamExplode2}.

\begin{figure}[t]
     \begin{subfigure}[c]{0.48\textwidth}
         \centering
         \includegraphics[width=\textwidth]{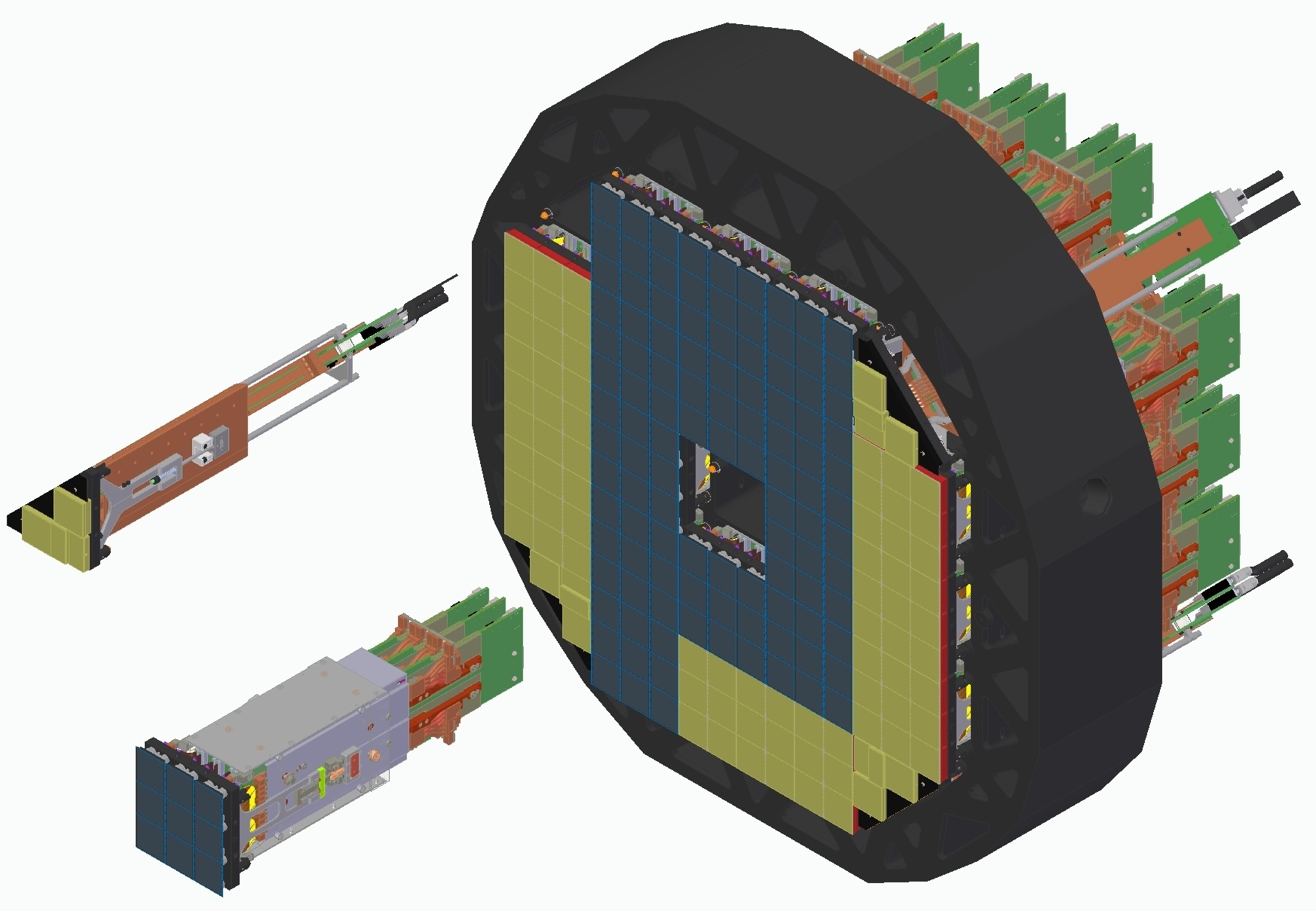}
         \label{fig:FocalPlaneExplodeSimplified}
     \end{subfigure}
     \hfill
     \begin{subfigure}[c]{0.48\textwidth}
         \centering
         \includegraphics[width=\textwidth]{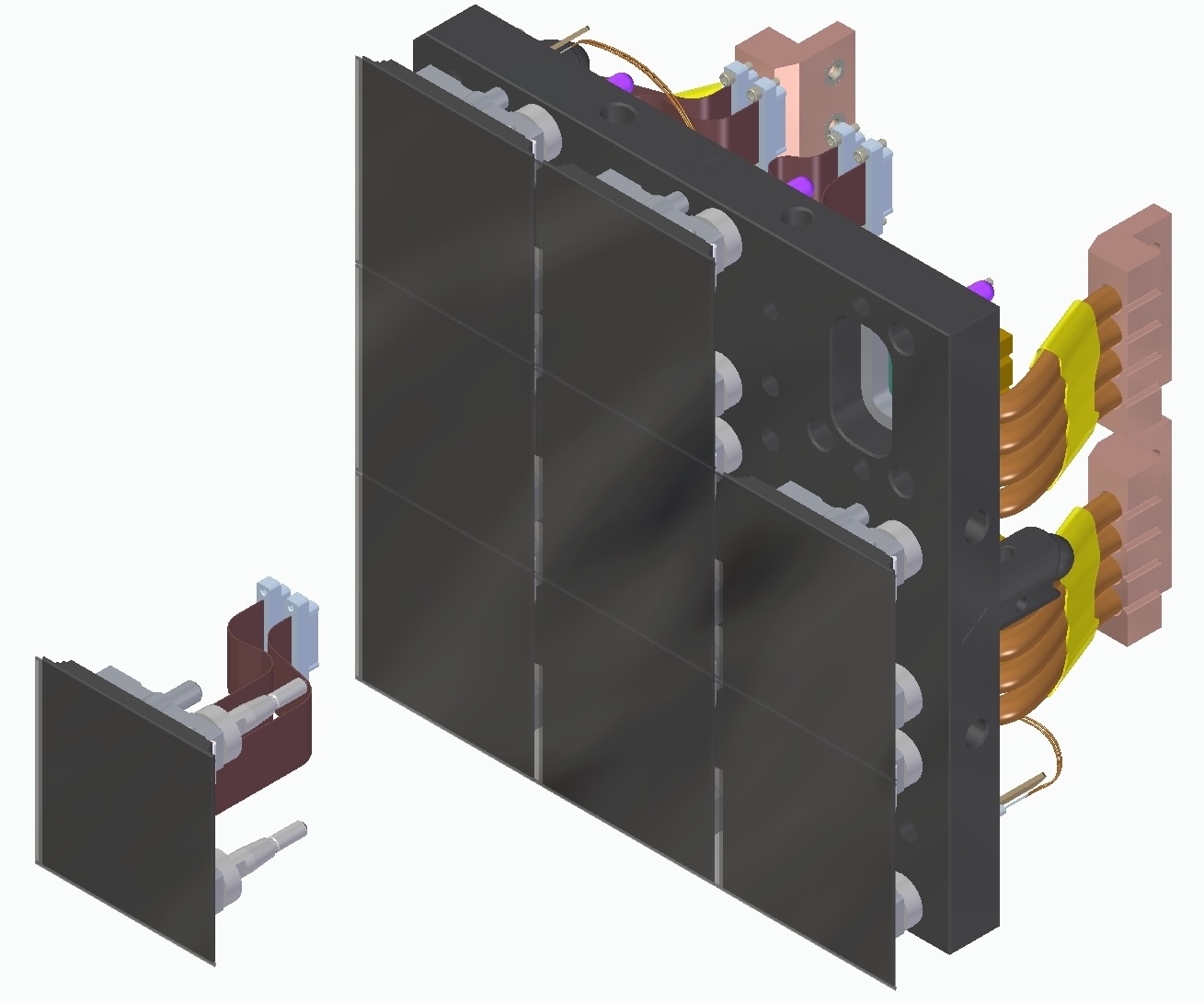}
         \label{fig:E2VCCDExplode2}
     \end{subfigure}

     \caption{Left: The focal-plane array of \ac{lsstcam}. Sensors produced by \ac{e2v} are shown in dark blue, while sensors produced by \ac{itl} are shown in yellow. One science and corner \ac{RTM} are extracted from the focal plane for comparison. Right: One $3\times3$ CCD array, contained in each science raft, with one sensor extracted for scale.}
     \label{fig:focal_plane_rtm}
\end{figure}

\begin{figure}[t]
         \centering
         \includegraphics[width=\textwidth]{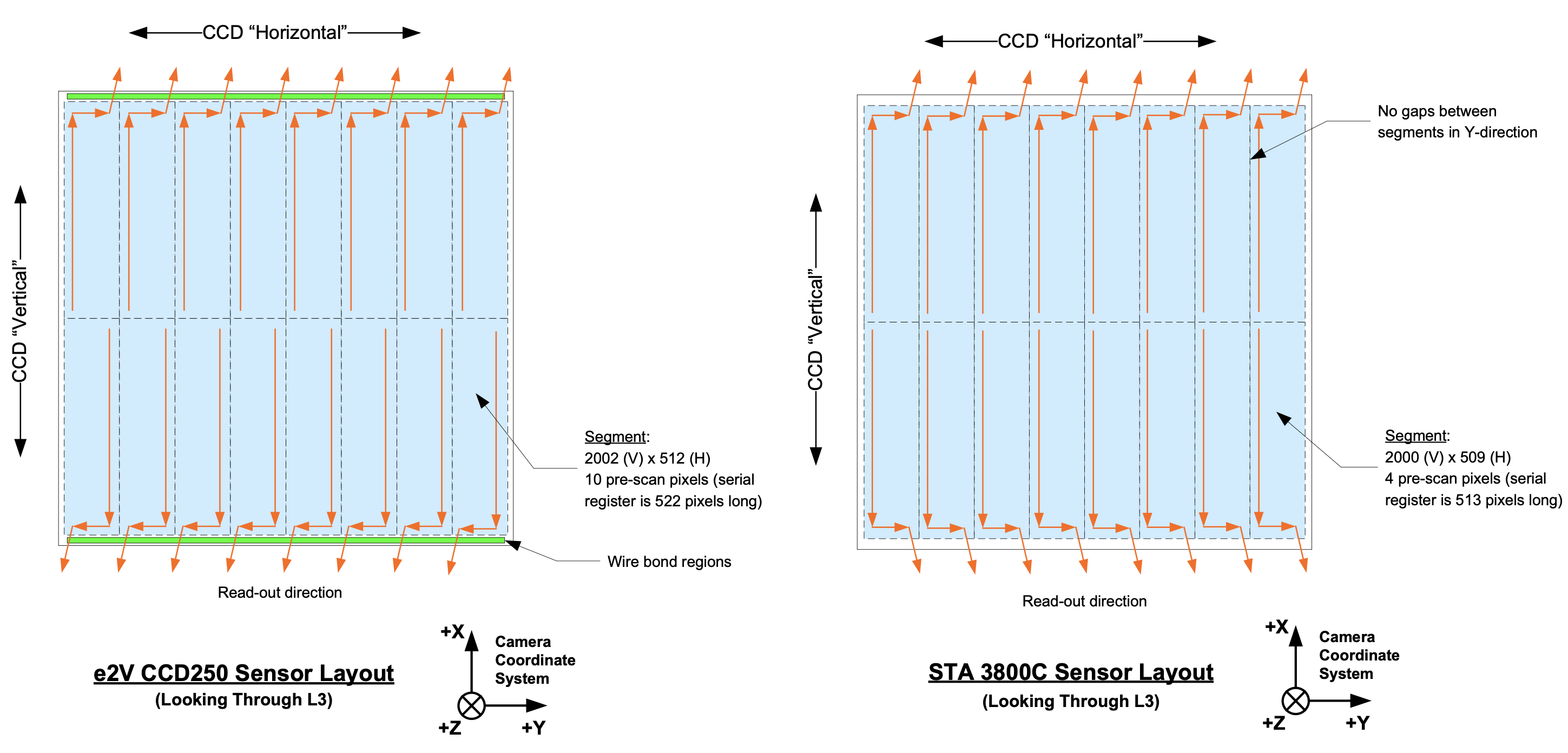}
         \label{fig:CCD_schematics}
     \caption{Diagram of the \ac{e2v} and \ac{itl} science \ac{ccds}. Orange lines indicate the readout direction. The angled orange lines indicate the location of the readout amplifier. \ac{e2v} \ac{ccds} have wire bond regions and a mid-line implant, while the \ac{itl} sensors do not.}
     \label{fig:composite_layout}
\end{figure}

The sensors used in the focal plane are \ac{ccds} from two manufacturers:  CCD250 from \ac{e2v} and STA-3800C from \ac{itl}. The science focal plane is composed of 189 \ac{ccds}. Science \ac{ccds} are arranged in $3\times3$ arrays with their readout electronics to form a single \ac{RTM}.\cite{CTN-001} The science focal plane has 21 \ac{RTM}s. Each \ac{RTM}s is composed of \ac{ccds} from a single vendor - \ac{e2v} or \ac{itl}. Thirteen \ac{e2v} \ac{RTM}s and 9 \ac{itl} \ac{RTM}s make the science focal plane. Each science CCD is $\sim4\text{k}\times4\text{k}$ pixels, read out in parallel via sixteen amplifiers, each responsible for a $\sim500\times2000$ pixel segment. The science focal plane is supplemented by 8 wavefront \ac{ccds} and 8 guider \ac{ccds} in the corners of the focal plane, each installed in one of 4 corner \ac{RTM}s. The corner \ac{ccds} are used for guiding and focusing \ac{lsstcam} through active feedback to the observatory. The guider \ac{ccds} are \ac{itl} STA-3800C CCDs, and the wavefront \ac{ccds} are \ac{itl} STA-4400B CCDs, each $\sim$ 4k $\times$ $\sim$ 2k.  On each corner raft the wavefront sensors are mounted adjacent to each other but with offsets $\pm2$\,mm along the optical axis to capture intra/extra-focal images \cite{2014SPIE.9147E..74R}.  They are read out in parallel using eight amplifiers, each responsible for $\sim500\times2000$ pixels. Diagrams showing the full focal plane layout as well as the the individual \ac{RTM} layout can be found in Figure \ref{fig:focal_plane_rtm}. Diagrams showing the readout of each science CCD type is shown in Figure \ref{fig:composite_layout}.

The pixel pitch of the \ac{lsstcam} \ac{ccds} is 10\,$\mu$m, corresponding to $0.2$ arcseconds per pixel in the Rubin optical system.  The \ac{ccds} are back illuminated with $100\mu\text{m}$ thick Si, for high sensitivity in red wavelengths. The \ac{ccds} are operated in fully depleted mode. The image format and operating voltages are slightly different for \ac{e2v} and \ac{itl} \ac{ccds}. The entire focal plane can be read out in parallel in 2 seconds but for improved read noise it has been optimized to 3 seconds.

\ac{ccds} were delivered to \ac{BNL}, where they were individually characterized and integrated into the science \ac{RTM}s.\cite{2018SPIE10702E..2CL} The science \ac{RTM}s were delivered to \ac{SLAC}, where they were re-tested before integration to the SiC grid inside the \ac{lsstcam} cryostat.\cite{2024SPIE13096E..1OL} Corner \ac{ccds} were received at \ac{SLAC} and assembled into the corner \ac{RTM}s, before testing in a small cryostat with a flat illuminator and tunable LED. The 25th and final \ac{RTM} was installed on January 14th, 2020, and was followed by the Camera integration. Short periods of electro-optical testing took place after integration of 2 and 9 \ac{RTM} modules. After the \ac{RTM}s were integrated, the focal plane was tested at \ac{SLAC} to verify performance before shipment to Cerro Pachón, Chile. Integration and testing of \ac{lsstcam} at \ac{SLAC} was completed in March 2024,\cite{2024SPIE13103E..0WU} and \ac{lsstcam} was shipped to Chile in May 2024.

In this paper, we present the characterization and performance of the LSSTCam CCD array from receipt on summit in May 2024 through early LSST operations at the Rubin Observatory. Section \ref{sec:run7_reverification} discusses the electro-optical re-verification of \ac{lsstcam} in the Rubin Observatory level 3 clean room. We discuss the integration to Rubin Observatory and the Simonyi Survey Telescope in section \ref{sec:camera_integration_to_Rubin}. In section \ref{sec:CCD_performance_on_sky}, we discuss the on-sky commissioning campaign of the \ac{lsstcam}, including the on-sky tests and optimizations to the focal plane. We discuss sensor features and anomalies in section \ref{sec:CCD_features_anomalies}. Finally, we discuss the current status of \ac{lsstcam} ahead of the start of the \ac{lsst} in section \ref{sec:conclusion}.

\section{Re-verification at Cerro Pachon}\label{sec:run7_reverification}
In May 2024, \ac{lsstcam} was loaded into a Boeing 747 airplane in San Francisco, CA, and flown to Arturo Merino Benítez International Airport in Santiago Chile. After receipt, \ac{lsstcam} was transported by truck more than $500\text{ km}$ from Santiago, Chile to the summit of Cerro Pachón at $2700\text{ m}$ in the Andes mountain range, where the NSF-DOE Vera C. Rubin Observatory was being constructed. After receipt on summit, \ac{lsstcam} was transferred to a support structure and rolled into the clean room, known as the White Room, on Level 3 of the Vera C. Rubin Observatory. After connecting lines for power, cooling, and verifying the integrity of the vacuum, \ac{lsstcam} was cooled down starting in late August 2024. Following cooldown, the seventh series of \ac{EOT} began, known as Run 7. This was the last testing and characterization opportunity for \ac{lsstcam} before installation on the \ac{TMA}, the mechanical structure that holds the Simonyi Survey telescope mirrors and \ac{lsstcam} once integrated. 

Run 7 was conducted from the end of September 2024 to the beginning of December 2024 to reverify \ac{lsstcam}'s performance and focal-plane optimization. Over the entire testing run, we collected 56,066 exposures, totaling 459.74TB during the three month period. All data was sent to the processing nodes at the \ac{S3DF}, the compute, storage, and network architecture designed to support Rubin Observatory and other experiments. The central activities and findings of the testing run are summarized here, with a detailed analysis of these and other features from the Run 7 period available in \texttt{SITCOMTN-148}.\cite{SITCOMTN-148}

\subsection{Electro-optical test stands}\label{subsec:EO_test_stands}

After receipt on summit, we re-characterized, optimized, and tested \ac{lsstcam} in the Level 3 clean room at Rubin Observatory. We call this electro-optical testing period \ac{EOT}. We used the \ac{CCOB} wide-beam as the primary electro-optical scene projector during \ac{EOT} \cite{2024SPIE13103E..0WU}.\footnote{In addition to the CCOB, we used a 4K projector (Epson LS11000 LCD) for spot projection in Run 7. This projector was first tested at \ac{SLAC} and arrived at Vera C. Rubin Observatory halfway through Run 7. The projector can produce spots on all 3216 amplifiers instead of just the 21 illuminated using the pinhole filter. Since the projector does not have fast illumination control, the \ac{lsstcam} main shutter was used, in comparison to pulsing the light source with the \ac{CCOB} Wide Beam. One observed downside is that the projector has a residual background level illumination outside of the spot regions. Additionally, the background illumination has time dependent structures which could not be easily subtracted. The resulting contrast between circular spots and the background was only about a factor of 6. Changing the spot shape to large rectangles increased the contrast ratio to 30. This contrast ratio was insufficient for high precision crosstalk and persistence tests, and thus did not provide data that contributed to \ac{lsstcam} optimizations during the \ac{EOT} campaign.}

\begin{figure}[b]
    \centering
    \includegraphics[width=0.95\linewidth]{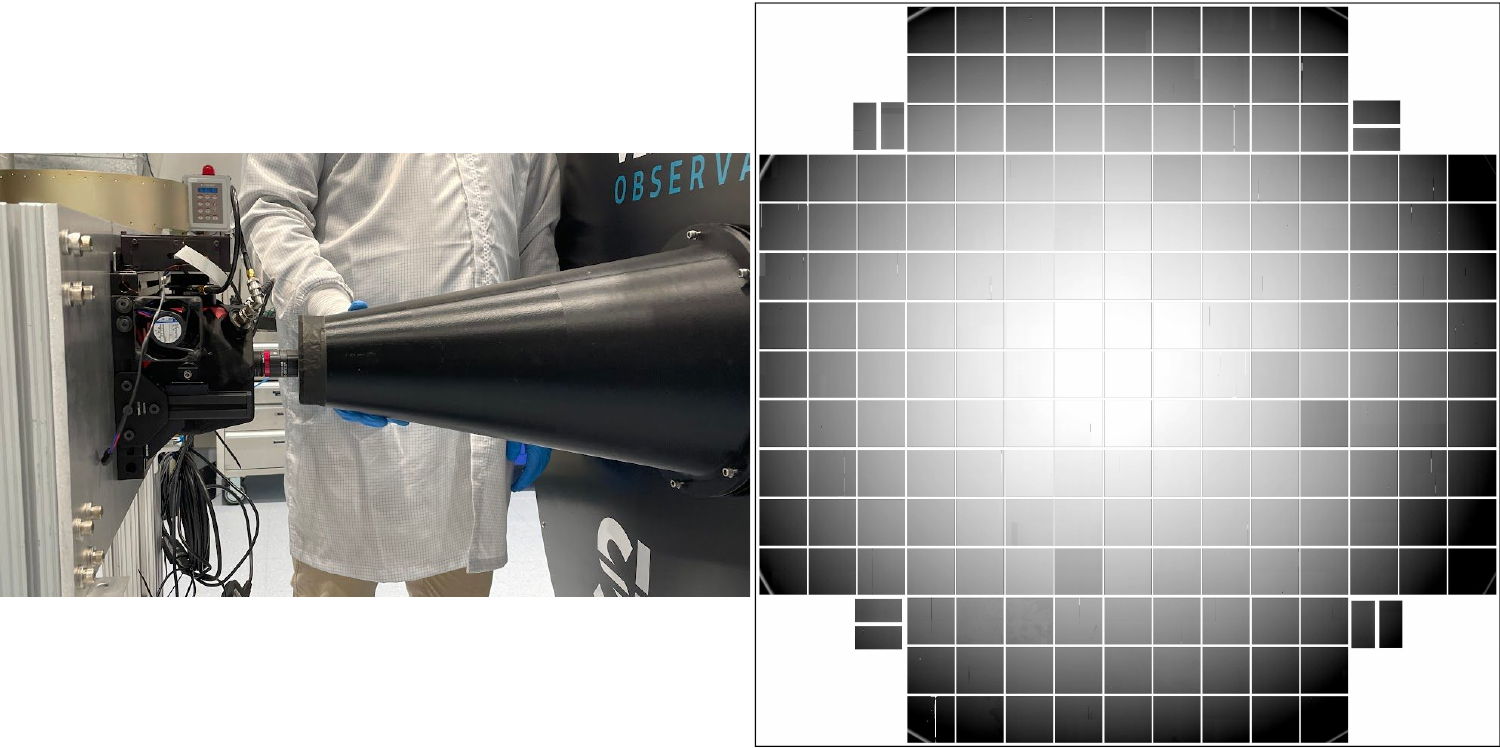}
    \caption{
    Left: The \ac{CCOB} wide beam projector during installation in the level three clean room of Rubin Observatory. 
    Right: The projected image from the \ac{CCOB} wide beam projector, illuminating the \ac{lsstcam} focal plane at $\lambda=750$nm. The illumination wavelength is controlled by the LEDs from the \ac{CCOB}.
    }
    \label{fig:ccob_4k}
\end{figure}

\ac{CCOB} wide-beam is designed to project a uniform monochromatic light source onto the entire \ac{lsstcam} focal plane, using the total exposure time to control the flux level. Light is produced by one of six LEDs, each corresponding to the central wavelength in one \ac{lsst} filter, controlled by a custom LED driver, and fed into an integrating sphere. The light from the integrating sphere is projected onto the focal plane with an ND filter (10\%) attached to a C-mount lens with an adjustable f/stop, shown in the left of Figure \ref{fig:ccob_4k}. 


For Run 7 we achieved a dark current of \textless0.1 ADU/sec with the shutter open and the \ac{CCOB} wide-beam installed through effective shrouding of the body of \ac{lsstcam} and the \ac{CCOB}. One difference between previous \ac{EOT} runs and Run 7 was that the f/stop of the lens was changed from 2.6 to 1.6, corresponding to a fully open aperture. We made this change to address two effects that were observed in previous \ac{EOT} runs; a time-dependent turbulent `weather' pattern from air circulation in the integrating sphere, and a static surface pattern from the inside integrating sphere, referred to as the \ac{cmb}, due to its similarity to the temperature fluctuations visible in the \ac{cmb}.\cite{2025JATIS..11a1205B} While changing the f/stop  did reduce the flux variations from the `weather' pattern, it also caused a much steeper illumination roll-off from the center to the edge of the focal plane. 

To retain uniform illumination across the focal plane while mitigating the 'weather' and 'CMB' effects, we installed a diffuser on the \ac{CCOB}-wide beam for Run 7. The diffuser is a $60^\circ$ diffusing angle sheet from Edmunds Optics. We found that the diffuser greatly reduces the `weather', eliminates the CMB pattern, and more uniformly illuminates the focal plane, with a penalty of decreasing the overall illumination by roughly 35\%. The diffuser was installed for all Run 7 characterization measurements using the \ac{CCOB}-wide beam.

At the start of the \ac{EOT} period, the g, r, y, pinhole, and empty filters were loaded in the \ac{lsstcam} filter carousel. The most common configuration used was the 'open' filter, an empty filter frame that passes all wavelengths of light. For chromatic studies, we relied on the LEDs in the \ac{CCOB} Wide projector. Another common setup used the pinhole filter to measure crosstalk and persistence. When combined with the \ac{CCOB} wide-beam, this illuminated a single amplifier of the central detector of each of the 21 \ac{RTM} modules.

\subsection{\ac{lsstcam} reverification}\label{subsec:summit reverification}

After the \ac{lsstcam} cryostat was cooled down to an operating temperature of $150$ K at the focal plane, it was paramount to reverify the performance of \ac{lsstcam} after the journey from \ac{SLAC} to Cerro Pachón. To reverify \ac{lsstcam}, we compared data acquired at the summit to data acquired at \ac{SLAC} under identical operating conditions, to identify any degradation in performance during shipment.

Two primary test sequences were used for summit re-verification and characterization:
\begin{itemize}
    \item \textbf{B protocols}: This sequence consists of the minimal set of camera acquisitions for \ac{EOT}, including bias images, dark images, \ac{flats} taken at various flux levels to construct a \ac{ptc}, \ac{flats} taken at constant flux levels, \ac{flats} taken with different LEDs, and a persistence dataset (a saturated flat, followed by several dark acquisitions to evaluate residual charge). This dataset allows us to measure a variety of sensor parameters, including:
    \begin{itemize}
        \item Dark current: Dark current is computed as the median value of a dark image after applying a gain correction, scaled by the amplifier gains and normalized by the integration time.
        \item Read noise: Read noise is computed as the standard deviation of the bias data for each CCD segment.
        \item Charge transfer inefficiency (CTI): CTI is measured using the extended pixel edge response (EPER) method, which measures CTI as the ratio between the overscan pixel signal and the product of the number of transfers in the overscan region and the signal in the last imaging pixel\cite{2020arXiv200103223S}.
        \item Sensor defects: we classify sensor defects as either bright or dark, and flag them as pixels and/or columns that are outliers relative to flat or dark images. For Run 7, we identify bright defects as any pixel that demonstrates a dark current greater than 5 e- s$^{-1}$, and identify dark defects as any pixel that is 20\% lower than the median flat field value.
        For on-sky commissioning, we identify bright defects as any pixel that demonstrates a dark current greater than 3 e-, and identify dark defects as any pixel that is 10\% lower than the median flat field value.
        \item Amplifier tearing: amplifier tearing appears as a signal deficiency at amplifier boundaries in \ac{e2v} sensors, accompanied by increased signal in adjacent columns.
        \item Persistence: this is a residual charge defect, with incomplete readout of charge from saturated pixels, allowing charge to 'persist' after a saturated exposure. We measure this signal by saturating the focal plane, and taking a series of dark images to measure if persistence is present.
    \end{itemize}
    \item \textbf{\ac{PTC}}: This sequence consists of a set of acquisitions of pairs of flat images taken at different flux levels. This flat acquisition sequence samples a wider range of flux levels at a higher density compared to the B protocol flat sequence, enabling more precise estimates of flat pair metrics including pixel covariances. This dataset allows us to measure a variety of sensor parameters, including:
    \begin{itemize}
        \item \ac{PTC} gain: \ac{PTC} gain is found by fitting the variance vs. mean from the PTC data following Astier et al.\cite{astier2019shape}
        \item \ac{PTC} turnoff: \ac{PTC} turnoff occurs when pixels become saturated and bloom, suppressing the variance. We compute the \ac{PTC} turnoff as the flux value where the variance of the PTC curve starts decreasing monotonically at higher flux levels.
        \item Brighter-fatter effect: the Brighter-fatter effect describes the pixel-pixel interations in response to changes in effective pixel area from the accumulated charges in the potential wells. In practice, the Brighter-fatter effect broadens the light profiles of bright sources. We measure the Brighter-fatter effect following the model described by Astier et al.\cite{astier2019shape} and implemented for \ac{lsstcam} analysis in Broughton et al.\cite{2024PASP..136d5003B}.
    \end{itemize}
\end{itemize}

All characterization data was acquired using these test sequences, and sent to \ac{S3DF} for processing. Raw \ac{lsstcam} data was processed using the \ac{cp_pipe} and \ac{eo_pipe}, used to produce calibrated images and derived electro-optical metrics, respectively.\cite{PSTN-019} The derived electro-optical metrics from \ac{eo_pipe} are the primary quantities used to characterize the \ac{lsstcam} focal plane sensors.

All camera performance metrics from the summit show close agreement with previous \ac{EOT} data acquired at \ac{SLAC}, as shown in Table \ref{tab:initRever:Table}. \ac{PTC} and full-well metrics were consistent, and no significant bright cosmetic defects had developed. Dark cosmetic defects were difficult to quantify due to sensor edge effects, though a comparison of previous run performance with a fixed picture frame showed that defects did not significantly change. Dark current and amplifier tearing showed improved performance compared to previous runs, while the persistence feature was still prominent in \ac{e2v} sensors.

\subsection{\ac{lsstcam} optimization}\label{subsec:summit optimization}

With the baseline condition of \ac{lsstcam} established, several aspects of optimization of sensor performance were considered.

\subsubsection{Persistence}\label{subsubsec:persistence_eot}

A well-documented sensor effect in \ac{e2v} sensors is persistence.\cite{DOHERTY,DMTN-276,2024SPIE13103E..0WU,2025JInst..20P7031P} Only the \ac{e2v} sensors have detectable persistence. A single saturated source leads to two separate persistence effects: at the location of saturation in subsequent images, and as a trail in the direction of readout in subsequent frames. While both contaminants are related to the same physical source, they have different decay rates, making image processing of persistence features an arduous task. 

We evaluated two different approaches to mitigate persistence: 1) Establishing a pinning condition where the holes form a thin layer at the front side of the CCD so that the excess charges recombine with these holes, and 2) Narrowing the parallel swing voltage during readout so that the charges accumulated in the silicon do not get close to the surface states responsible for trapping. Following consultation with Teledyne \ac{e2v}, we decided to narrow the parallel swing, as the parallel low voltage would need to be lowered to $-$7.0\,V or lower to establish the pinning condition. With parallel low voltages at this level, the measured current flow increased, which in turn Teledyne \ac{e2v} advised would increase the risk of damaging the sensor.

\begin{figure}[t]
    \centering
    \includegraphics[width=0.495\linewidth]{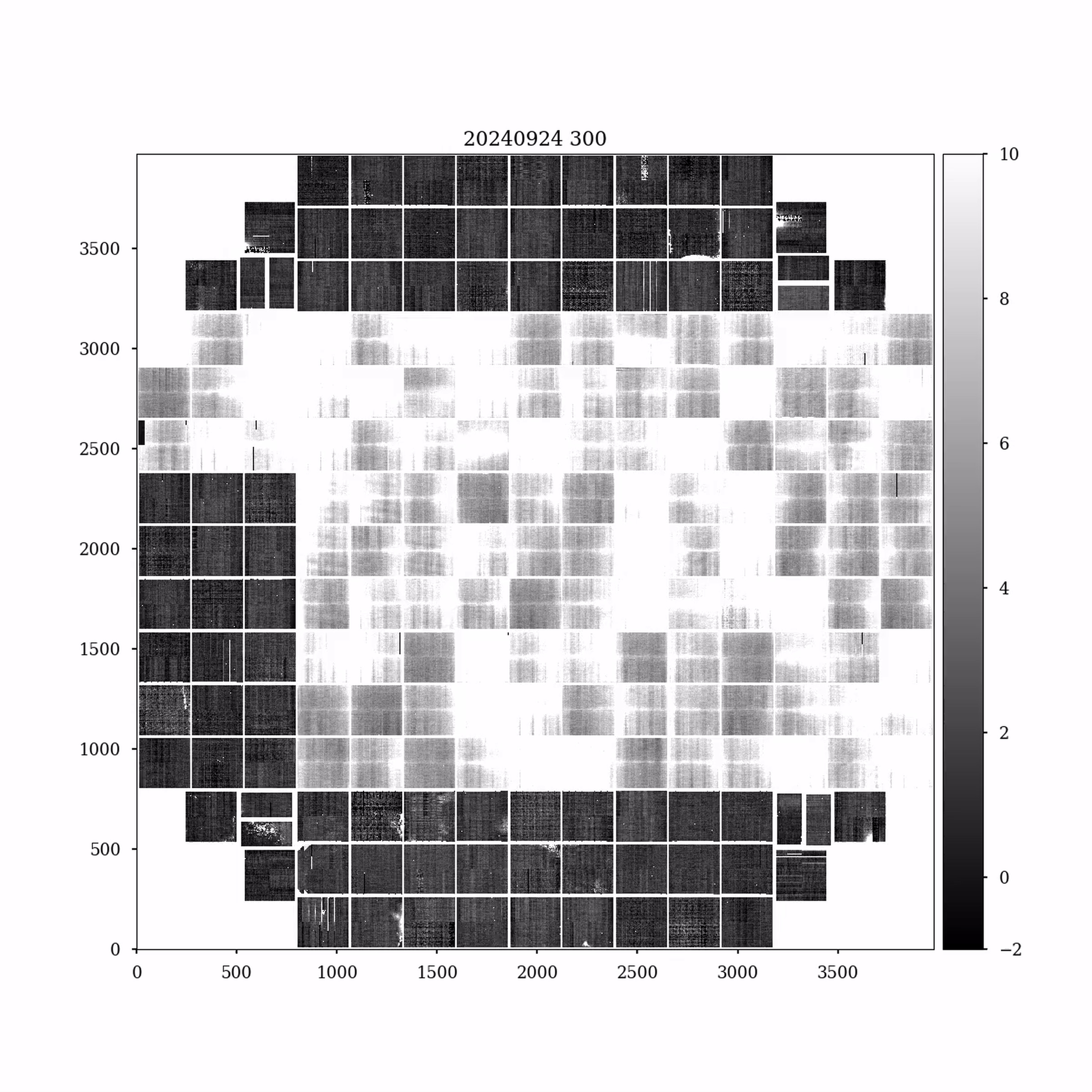}
    \includegraphics[width=0.495\linewidth]{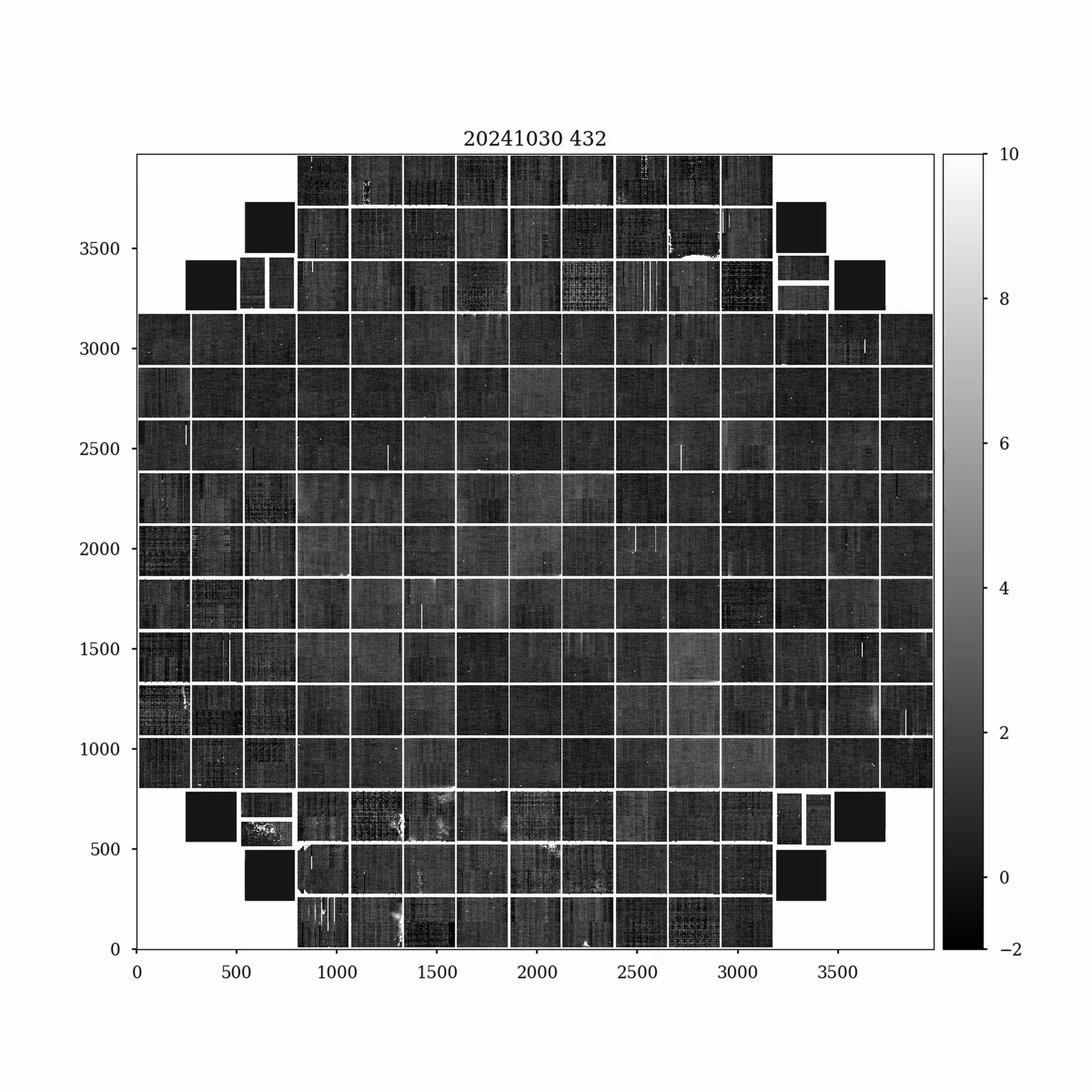}
    \caption{The persistence effect in \ac{e2v} sensors. The test procedure used to quantify persistence was to overexpose the focal-plane and obtain one exposure before closing the \ac{lsstcam} shutter and obtaining multiple dark images. Exposures shown are the first dark image taken after overexposure with \ac{lsstcam} shutter closed. Left: The residual charge present in \ac{e2v} sensors before optimization, with the residual charge signal of $8.3$ e- visible in the central 117 \ac{e2v} \ac{ccds}. Right: The first dark exposure taken from after lowering the parallel, with the residual charge signal of $0.6$ e- in the \ac{e2v} \ac{ccds}.
    }
    \label{fig:persistence_mitigation}
\end{figure}

Testing performed at \ac{UCD} on a spare \ac{e2v} CCD confirmed that the amplitude of the persistence decreases as the parallel swing voltage is decreased for \ac{e2v} sensors \cite{2014SPIE.9154E..15T}. Based on these results, we performed tests with different operating voltages, characterized by the parallel swing voltage (9.3\,V, 8.8\,V, 8.4\,V, and 8.0\,V), and  we observed a similar decrease in persistence to the sub-ADU level. The residual charge signal in images is shown in Figure \ref{fig:persistence_mitigation}.

However, these voltage changes lead to two consequences:
\begin{enumerate}
    \item \textbf{Lower full well capacity}: With the new parallel voltage configuration, we observe a 23\% reduction in \ac{PTC} turnoff. A reduction of full-well capacity is expected, as the observation of persistence indicates that the sensor is in a surface full-well condition, not an optimal or blooming full-well condition. To achieve either alternative, constraining the parallel swing and consequently the full well, is expected to be needed.
    \item \textbf{Increased brighter-fatter amplitude}: We observe a 5--19\% increase in the strength of the brighter-fatter effect, with the difference in strength of brighter-fatter signal measured relative to the Run 7 reverification operating conditions. However, the brighter-fatter amplitude does not become significantly more anisotropic: the ratio of parallel to serial nearest-neighbor correlations increases only from 3.43 to 3.54, so the reduction of parallel swing does not risk increasing systematic uncertainty of the PSF ellipticity.
\end{enumerate}

\subsubsection{Sequencer optimization}\label{subsubsec:R7_seqOpt}

The clocking sequence for \ac{lsstcam} \ac{ccds} is defined via ASCII formatted sequencer files. This configuration defines the sensor clearing, the image readout sequence, and the clock timing durations. Several tests with different sequencer configurations were carried out during Run 7.

\begin{itemize}
    \item \textbf{Clear activity}: Different sensor clear procedures were designed to remove residual charge from pixel rows near the serial register. Based on testing of the different procedures, the \texttt{NopSf} clear method was determined to remove all residual electrons near the serial register in the shortest clocking time, and therefore it was adopted as the new default. Previously, we kept all clock phases high for a fast clear of the charges along the serial register. \texttt{NopSf} implements a clear where the serial register has serial clock 1 \& 2 high, and serial clock 3 low when the parallel clock 1 moves the charges to the serial register from the image region. 
    \item \textbf{\ac{RG} output}: During parallel transfer, the CCD amplifier is commonly protected from large signal injection associated with the parallel clock swing by activating the \ac{RG} of the amplifier. The default sequencer configuration had the \ac{RG} output disabled for \ac{e2v} sensors and enabled for \ac{itl} sensors following a 2024 study.\cite{2024SPIE13103E..0WU} Obtaining images with the \ac{RG} output enabled in \ac{e2v} sensors, we observed that the amplitude of bias structure variations improved, however the stability of the electronic bias structure over time did not improve. Despite this residual time dependence, we enabled the \ac{RG} output by default for \ac{e2v} sensors following this testing.
    \item \textbf{Idle Flush activity}: Idle Flush is a configuration in the sequencer file which enables the sequencer output to run while in the IDLE state. While Idle Flush was enabled to address structure visible in bias frames, we identified that Idle Flush exacerbates the amplifier tearing signal\cite{2020arXiv200209439J}. Additionally, the default Idle Flush function used a regular readout operation all  the time, which has a significant thermal impact because it continuously operates the Analog-to-Digital converters at their high rate. Based on these observations we disabled Idle Flush, thus improving the overall thermal stability and minimizing the amplifier tearing signal.
    \item \textbf{Phase overlap during parallel transfer}: \ac{e2v} sensors feature four parallel clock phases. To improve the uniformity of the full well across a sensor, overlapping two phases during each time slice of the parallel transfer was introduced as the new default.
\end{itemize}

\begin{table}[b]
\centering
\resizebox{\textwidth}{!}{ 
\begin{tabular}{|c|c|c|c|c|c|c|c|}
\hline
 &  & \multicolumn{3}{c|}{e2v}                   & \multicolumn{3}{c|}{ITL}                    \\ \cline{3-8}
Parameter [unit] & Specification & \makecell{\ac{SLAC}\\testing} & \makecell{Summit \\ reverification} & \makecell{Summit final \\ characterization}     & \makecell{\ac{SLAC}\\testing} & \makecell{Summit \\ reverification} & \makecell{Summit final \\ characterization} \\ \hline
Dark current {[}e- pix$^{-1}$ s$^{-1}${]}        & None               & 0.055 & 0.025& 0.023 & 0.046 & 0.021 &     0.021      \\ \hline
Read noise {[}e-{]}            & $<9$ e-               & 5.30    &   5.32  &   5.40   & 6.20    &  6.26   &     6.21     \\ \hline
Serial CTI {[}\%{]}                & $<5\times 10^{-4}$               & 3.7$\times 10^{-5}$ & 1.1$\times 10^{-5}$&7.3$\times10^{-6}$  & 1.2$\times 10^{-4}$ & 1.7$\times 10^{-4}$ &    1.5$\times10^{-4}$     \\ \hline
Parallel CTI {[}\%{]}              & $<3\times 10^{-4}$               & 1.2$\times 10^{-5}$ & 1.2$\times 10^{-5}$&  1.1$\times10^{-5}$ & 3.4$\times 10^{-7}$ & -4.8$\times 10^{-6}$ &    1.2$\times10^{-6}$      \\ \hline
PTC turnoff {[}e-{]}               & $>90\text{k}$ e-  & 126,000  & 133,000  & 103,000  & 117,000  & 129,000 &         129,000     \\ \hline
PTC Gain {[}e- / ADU{]}            & None               & 1.48    & 1.48    &   1.51  & 1.67    & 1.68  &          1.68      \\ \hline
PTC $|a_{00}|$ [$10^{-6} \text{ e-}^{-2}$]   & None               & 3.09 & 3.09&     3.49     & 1.71 & 1.70 &        1.70     \\ \hline
Bright defects {[}count{]}         & None               & 0         & 0         &  0  & 0          & 0 &             0   \\ \hline
Dark defects {[}count{]}           & $<20\text{k}$ per amp & 4 & 3 & 3    & 9 & 8 &        7      \\ \hline
Persistence {[}ADU{]}              & None               & 5.67    & 5.64    & 0.40 & 0.48   & 0.42    &       0.33      \\ \hline
\end{tabular}
}

\caption{Comparison of the median values of different parameters, obtained during testing in the IR2 clean room at \ac{SLAC}, during initial re-verification testing at Rubin Observatory, and during the final characterization after \ac{EOT} optimization. Results are shown per detector type. Only science detectors are considered in this table. $a_{00}$ represents the relative strength (to first order) of the brighter-fatter effect, and is negative by sign convention of Astier et. al.\cite{astier2019shape}}\label{tab:initRever:Table}
\end{table}

\subsection{\ac{lsstcam} characterization}\label{subsec:summit characterization}

Following the \ac{lsstcam} optimization, a final characterization was performed before installation on the Rubin Observatory \ac{TMA}. This characterization was performed with the operating configuration established during the \ac{lsstcam} optimization, using the same test sequences as reverification and analyzed using the same image-processing pipelines.

Comparison of the initial and final Run 7 \ac{lsstcam} performance at the Rubin Observatory highlights the consistency of the values of many electro-optical metrics, as shown in Table \ref{tab:initRever:Table}. Serial and parallel CTI measurements are highly consistent between runs, showing slight improvements in the final operating configuration. Dark current measurements remain stable across the focal plane, with notable improvements in certain rafts due to light leak mitigation. Bright and dark defects show close agreement between runs, with no measured development of defects during final characterization. Read noise remains consistent with previous measurements for both \ac{e2v} and ITL sensors.

A major achievement from Run 7 is the substantial decrease in persistence signal for \ac{e2v} sensors, due to the lower parallel swing during readout. The trade-off from the change in parallel swing is a decrease in the \ac{PTC} turnoff, and an increase in the strength of the brighter fatter effect for \ac{e2v} sensors. The performance of the \ac{itl} sensors did not experience a significant change, as the operating voltages were not modified for these sensors.
\section{Integration to Rubin Observatory}\label{sec:camera_integration_to_Rubin}





The transition of \ac{lsstcam} from a standalone instrument to its integrated state within the Rubin Observatory marked a critical phase in the commissioning timeline. Following the conclusion of the final \ac{EOT} phase on the summit, \ac{lsstcam} underwent an orchestrated integration process to prepare for installation on the Simonyi Survey Telescope.

\subsection{Thermal Cycle and Integration to the Simonyi Survey Telescope}\label{subsec:Thermal Cycle}

Following the completion of \ac{EOT}, the cryostat was brought to ambient temperature in December 2024. This warm-up was conducted under strict temperature controls ($\Delta T<25 \text{K}$\textdegree per hour) to mitigate mechanical stress to the \ac{REB}s, cryo plate, and the cryostat grid. The warm-up was a prerequisite for the mechanical integration of the camera to the \ac{TEA}, the structure that interfaces \ac{lsstcam} with the Simonyi Survey Telescope. 

After the cryostat of \ac{lsstcam} was brought to ambient temperature, \ac{lsstcam} was moved from the level 3 clean room of Rubin Observatory to the high bay, where \ac{lsstcam} was lifted by crane and integrated with the \ac{TEA}. Once integrated to the \ac{TEA}, testing on the rotator, hexapods, and VIP line installation could proceed before moving \ac{lsstcam} to level 7 of the Rubin Observatory. VIP lines, or vacuum insulated pipe lines, inner line and an outer jacket. The inner line carries the cold fluid, while a vacuum space between 
Consist of an inner pipe to transport cold fluid from the chiller to the REB electronics, and an outer pipe to provide insulation that minimizes heat transfer.

Transportation from level 3 to level 7 of Rubin Observatory took place over two days, and was facilitated with an industrial elevator. On the first day of this operation, \ac{lsstcam} was transported from the high bay of level 3 to the industrial elevator, transported to level 7, and removed from the elevator to the platform adjacent to the \ac{TMA}. On the second day, \ac{lsstcam} was lifted using a crane inside the Rubin Observatory dome and precisely positioned along the optical axis of Rubin Observatory, interfacing with the interior of the secondary mirror. This was accomplished through a coordinated lift with the crane inside the Rubin Observatory dome, skilled operators angling \ac{lsstcam} into alignment to the telescope from the level 7 floor, and skilled operators on the telescope platform ensuring that \ac{lsstcam} was properly aligned inside the secondary mirror.

After a successful lift and mechanical integration of \ac{lsstcam} into the Simonyi Survey Telescope, work began to route the cooling, vacuum, and power lines through the telescope mount and to \ac{lsstcam}. For further details on the integration of \ac{lsstcam} to Rubin Observatory and the Simonyi Survey Telescope, see \cite{FanningsPaper}.

\subsection{Restoring the Focal Plane Array after Integration to the Simonyi Survey Telescope}\label{subsec:ccdShort}

In March 2025, we re-established power to \ac{lsstcam} in the integrated telescope environment. With vacuum and cooling lines reinstalled and power re-established, the cryostat was cooled down, and the cryoplate reached its operating temperature of $150$ K.

Shortly after the stable cryostat condition was established, the \ac{ccds} were powered back on for the first time since December 2024. During power on and associated testing, two issues arose;

\begin{itemize}
    \item Following the normal CCD power on procedure, an electrical shorts test was performed before any CCD power cycle. During the first round of short tests, all sensors were nominal, with the exception of one science sensor (referred to by its focal plane position as R20\_S21). Review of telemetry found that the impediment to turn-on was caused by a faulty reading in the \ac{adc} of the reset-drain voltage line of R20\_S21. The ADC measurement was nominal during the Run 7 \ac{EOT} campaign, read at the negative end of the \ac{adc} range when in emulation mode before integration to the Simonyi Survey Telescope, and read at the positive end of the \ac{adc} range when in operation after integration to the Simonyi Survey Telescope. This reading was outside of the pre-defined safe limits for turn on, though was deemed to be nonphysical.

    We assessed that the sensor remained operable and did not pose a risk to the focal plane as a whole. Development began on modifying the short test and CCD power on procedure to bypass the faulty reading while maintaining the same safety standard for the affected sensor. Due to the sensor readout multiplexing across three \ac{ccds}, the sensors on the affected \ac{REB} remained off for the remainder of \ac{lsstcam} first photon preparations.

    \item After the Simonyi Survey Telescope was moved from zenith to horizon for the first time with \ac{lsstcam} integrated, another short test was performed ahead of sensor turn on. During the short test, an electrical short was triggered involving one of the science \ac{ccds} (referred to by its focal plane position as R30\_S12). Testing and telemetry review pinpointed the issue to an overcurrent in the output drain current during power on. Normal procedure in this circumstance was to power cycle the \ac{REB} and try the shorts test again. After a power cycle and another short test, another failure to power on was triggered due to high output drain current, accompanied by a small spike in the vacuum pressure. After the spike in vacuum pressure, we powered off all \ac{ccds} plus the \ac{REB} in question.
    
    Telemetry review and expert discussion concluded that the high output drain current is consistent with a short to ground on the output drain line to the CCD. The pads on the \ac{e2v} CCD adjacent to the output drain line are substrate ground. A short across those lines could cause the high current, and given the 200mA at 22V for a few seconds (the current and voltage driven during the short test), could blow out the 100 $\Omega$ resistor in front of each output drain line. If true, it is possible that only one amplifier of the CCD would be affected. If the resistor failed to blow out, and left a low resistance connection, then the whole CCD could be inoperable. 
    
    For the safety of the focal plane array, the three \ac{ccds} on the affected \ac{REB} were powered off while firmware development began, aiming to safely read out the other two \ac{ccds} integrated to the same \ac{REB}. For the acquisition of first photon, all \ac{ccds} on R30/Reb1 remained powered off.
\end{itemize}

\subsection{Pre-First Photon Focal Plane Status}\label{subsec:pre-first-photon}

By the eve of the first photon in April 15th, 2025, the \ac{lsstcam} focal plane reached a state of operational readiness. Of the 189 science CCDs, 183 were fully functional.  The vacuum system maintained a stable pressure ($\sim10^{-7}$ Torr), and the cryoplate temperature was actively regulated at $150\pm0.3$ K. Data quality verification exposures were taken inside the dome in a dark environment, including bias and dark current measurements, confirmed that the operation of the remaining \ac{ccds} was not impacted by the electrical short or other integration activities. The \ac{lsstcam} focal plane array was thus primed to capture its first high-resolution images of the southern sky, marking the beginning of the most ambitious wide-area galaxy survey in history.
\section{Commissioning the LSSTCam CCD array on sky}\label{sec:CCD_performance_on_sky}
On April 15, 2025, Rubin Observatory acquired the first on-sky exposures with \ac{lsstcam}, a milestone referred to as first photon. This exposure marked the first time the full focal-plane array was exposed to the sky, and the first exposure taken with the fully-integrated Rubin optical system and \ac{lsstcam}. In the fully integrated Rubin optical sensors, some edge CCD sensors are partially vignetted due to the 3.5$\deg$ diameter FOV inscribing a circle that extends to the diameter of the center row and column of \ac{ccds}. Once \ac{lsstcam} was on-sky, several tests were devised to evaluate the resilience of the focal plane array and optimize readout performance. Additionally, we focused on restoring operation to the \ac{ccds} that were turned off due to faulty \ac{adc} readings and an electrical short (see section \ref{subsec:ccdShort}).

\subsection{Focal Plane Performance Under High Flux}\label{subsec:BrightStarTests}

\ac{e2v} recommended to limit the current flow between the front and the back side to prevent a breakdown in the silicon substrate, which was subsequently adopted as a safety limit during operations of 120$\mu$A. The current flow depends on the charge accumulation, which raises a question on how bright a star can be safely observed. Addressing this question motivated dedicated on-sky tests during \ac{lsstcam} commissioning. Note that the automated shutdown of the high voltage across the silicon has been implemented, which is triggered by the current limit at $\sim$ 120$\mu$A per CCD to shutdown, which has a margin to the vendor-imposed limit of 200$\mu$A.


\begin{figure}[b]
    \centering
    \includegraphics[width=0.49\linewidth]{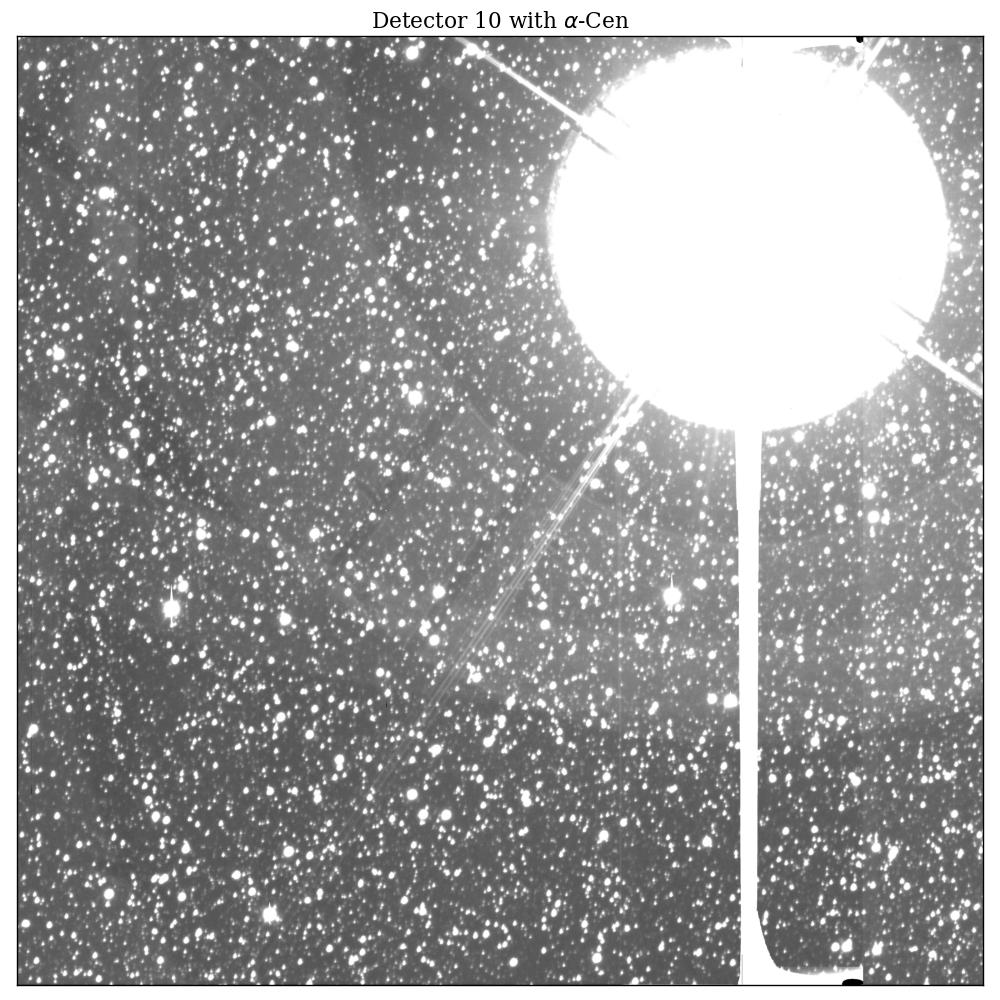}
    \includegraphics[width=0.49\linewidth]{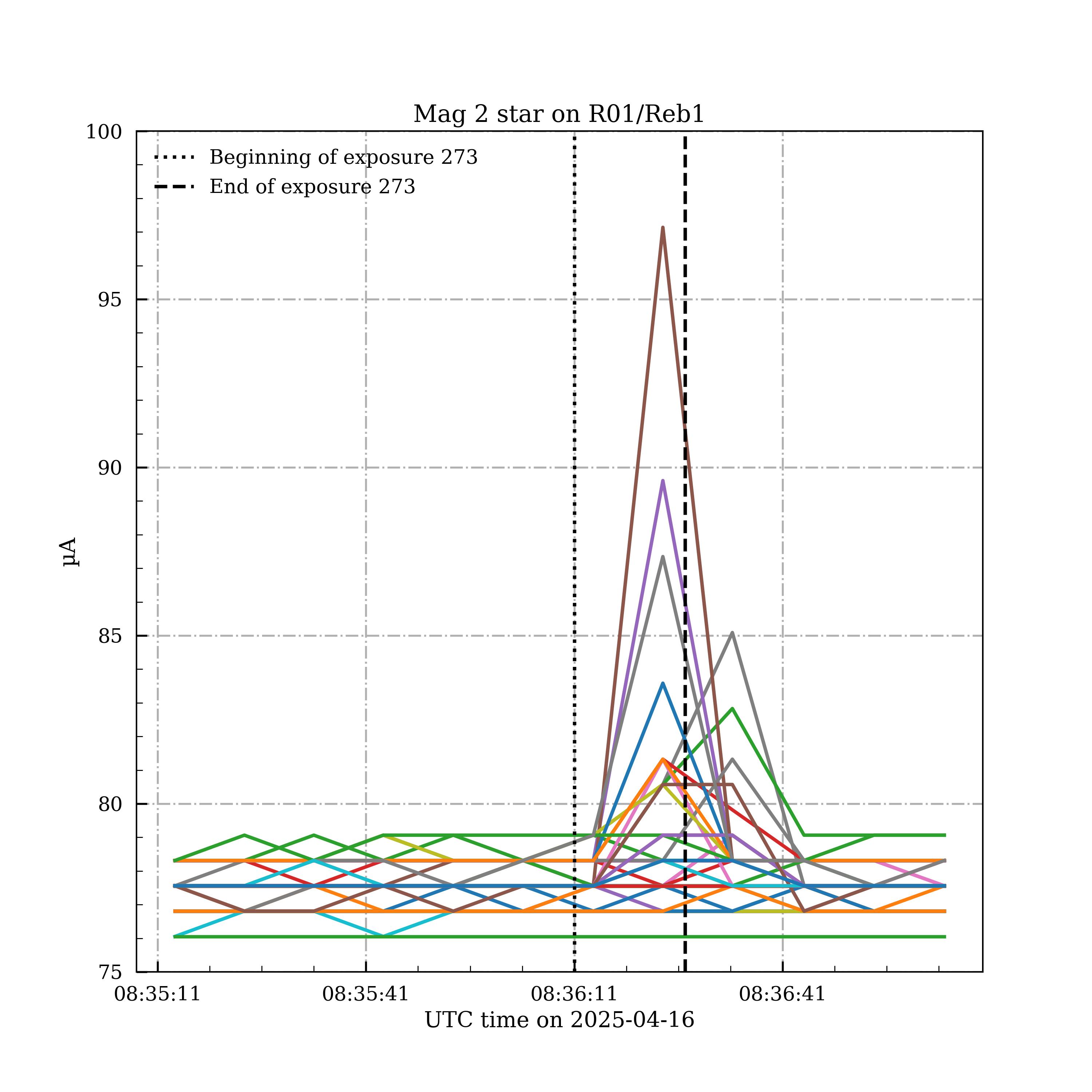}
    \caption{Left: The image of $\alpha$-Cen placed onto a CCD with a documented sensitivity to high flux exposures. Shown exposure, number 273 from that night, demonstrated the highest bias current fluctuation in response to high flux. Right: The largest bias current excursion observed during the bright star test, 97$\mu$A, from exposure 273 on the night of April 15, 2025. The vertical dashed lines indicate the start and end of the exposure. The solid lines indicate the measured current on each REB across the focal plane, with the largest current excursion observed on the REB with the bright star located on it.}
    \label{fig:bright_star_current}
\end{figure}


We devised a test to evaluate the sensitivity to bright sources, where we acquired a sequence of on-sky exposures with bright stars in the field-of-view and placed on \ac{REB}s that had tripped in response to high-flux exposures during the course of \ac{EOT} \cite{SITCOMTN-148}. The selected stars for testing have approximate apparent magnitudes of 0 (Alpha Centauri), 1 (tet Cen), and 2 (* gam Crv), and the test was executed in ascending order of stellar flux. 







For each exposure and REB, the time-series analysis focused on identifying sudden current excursions coincident with bright star illumination, sustained offsets relative to pre-exposure baselines, and differences in behavior between rafts or stellar magnitudes. The time of exposure start and end was used as the reference for identifying correlated behavior. 

Across all examined exposures, stellar magnitudes, and REBs, the bias-current telemetry remains within the acceptable safety limit of 120\,µA (for the three sensors including the baseline of 80\,µA). No current excursions above 97\,µA are observed (see Figure \ref{fig:bright_star_current}). No discontinuities, or systematic trends were observed during or after the bright-star exposures. This holds across the entire set of rafts tested, indicating that bright stars does not pose a risk to REBs that have shown previous sensitivity to high flux.



We conclude that bright star illumination at the tested magnitudes does not induce significant perturbations in \ac{REB} bias currents that could trigger an automated shutdown of the \ac{ccds}. This test provided confidence that \ac{lsstcam} electronics can operate consistently in the presence of bright optical sources up to magnitude 0, and that such sources do not pose a risk to sensor safety.


\subsection{Recovery of Non-Operational \ac{ccds}}\label{subsec:CCD_recovery}

As noted in section \ref{subsec:ccdShort}, six science \ac{ccds} were powered off for the acquisition of first photon. Following first photon acquisition, work began to address the two distinct problems preventing operation of the remaining \ac{ccds}.

To address the three \ac{ccds} associated with an erroneous \ac{adc} reading, modifications to the short test and power-on sequence were made to ensure that the faulty reading was bypassed without compromising the standard safety checks on the other voltage lines to that CCD. After testing on a spare \ac{RTM} at \ac{SLAC}, the firmware change was implemented to \ac{lsstcam} on April 24th, 2025, the affected REB was powered on, and successful operation of the three affected \ac{ccds} was confirmed with on-sky exposures that night. No adverse performance in those \ac{ccds} was observed in comparison to the \ac{EOT} campaign.

To address the three \ac{ccds} associated with an electrical short, significant modifications to the \ac{REB} firmware were made, and significant testing was conducted at the \ac{lsst} optical beam simulator at \ac{UCD}\cite{2014SPIE.9154E..15T} before deployment on sky. The firmware changes were to implement safety conditions to prevent hvBias turn on if measurements on certain voltage lines are too low (reset-drain, guard drain, output drain). Additional safety measures were implemented to the power-on procedures to verify correct operation of \ac{ccds}. Testing of the new firmware was conducted at \ac{UCD}, where \ac{e2v} \ac{ccds} were powered on and output-drain voltages were set to 0.15V. Before taking any images, the output-drain current, output-drain voltage, power supply voltage, and power supply current were compared to historical values. After verifying that these measurements were nominal, a series of bias images was acquired to verify operation while monitoring the power supply voltage and current. This image acquisition sequence was repeated with different values of the power-supply voltage, and found that the minimum measured output-drain voltage with the power-supply voltage oﬀ is 2.01V. As the power supply voltage increase, we observed that the output drain voltage and current decreased, along with the per-segment current. No anomalous fluctuations were observed in the power supply current or voltage during bias image acquisition. As a result of these tests, the firmware was deemed safe to deploy to \ac{lsstcam}. After testing was completed and a notice-to-proceed was issued, the firmware was loaded to \ac{lsstcam} on June 24th, 2025, and successful operation of the two operable \ac{ccds} was confirmed with on-sky exposures on June 25th, 2025. No adverse performance in those \ac{ccds} was observed in comparison to the \ac{EOT} campaign. We anticipate that the affected CCD will remain powered off for the foreseeable future.

After recovering the \ac{ccds} that were powered off during early operations, 188/189 science \ac{ccds} are operational in the \ac{lsstcam} focal plane.

\subsection{Sequencer and Idle Clear Optimizations}\label{subsec:onsky_sequencer_idleClear_optimizations}

Table \ref{tab:OnSkyPerformance:Table} shows the changes in key performance metrics of \ac{lsstcam} from the \ac{EOT} period to on-sky commissioning. While most metrics remain the same, read noise improved substantially for \ac{e2v} sensors due to optimizations of the \ac{lsstcam} sequencer. 

The Run 7 \ac{EOT} characterization results were obtained using a readout configuration defined as an ascii formatted sequencer text file. Following the Run 7 testing campaign, additional optimization targets were defined, focused on eliminating long-range correlations and reducing readout noise. Both of these optimizations were achieved through iterative improvement to the \ac{lsstcam} sequencer. 

\begin{table}[b]
\centering
\resizebox{\textwidth}{!}{ 
\begin{tabular}{|l|c|c|c|c|c|}
\hline
Parameter [unit] & Specification & \multicolumn{2}{c|}{e2v}                   & \multicolumn{2}{c|}{ITL}                    \\ \cline{3-6}
   &                                & \makecell{EO Testing \\ Characterization} & \makecell{On-telescope \\ Performance}    &  \makecell{EO Testing \\ Characterization} & \makecell{On-telescope \\ Performance} \\ \hline
Dark current {[}e- pix$^{-1}$ s$^{-1}${]}        & None   & 0.023 & 0.015  &    0.021 &    0.022  \\ \hline
Read noise {[}e-{]}            & $<9$ e-               &    5.40   & 4.96 &     6.21 &  6.32    \\ \hline
Serial CTI {[}\%{]}                & $<5\times 10^{-4}$               &$7.3\times10^{-6}$  & $3.1\times10^{-5}$ &    1.5$\times10^{-4}$ & 1.3$\times10^{-4}$    \\ \hline
PTC turnoff {[}e-{]}               & $>90\text{k}$ e-   & 103,000 & 102,000 &         129,000 & 120,000   \\ \hline
PTC Gain {[}e- / ADU{]}            & None               &  1.51  & 1.47 &          1.68 &    1.62  \\ \hline
PTC $|a_{00}|$ [$10^{-6} \text{ e-}^{-2}$]   & None               &      3.49     & 3.35 &        1.70 &   1.88  \\ \hline
\end{tabular}}
\caption{Comparison of the median values of different parameters, obtained during final re-verification testing at Rubin Observatory, and after on-sky testing and implementation of readout timing optimizations. Results are shown per detector type. Only science detectors are considered in this table. Notably, read noise for \ac{e2v} sensors decreased $\sim0.5$ e- due to the updated three second sequencer described in section \ref{subsubsec:R7_seqOpt}.}\label{tab:OnSkyPerformance:Table}
\end{table}

We tested series of modifications to the timing of the readout sequence on a spare \ac{RTM} at \ac{UCD}, with a focus on reducing long range correlated noise, which manifests as row-wise signals in bias images. After iterative testing on \ac{lsstcam}, a 20 ns delay to the reset-gate clock was selected as the best option, and added to the \ac{e2v} readout sequence to mitigate correlated noise. A similar modification was tested on \ac{itl} sensors, which had no effect on sensors that exhibited correlated noise. 

Additional modifications to the readout sequence were made in response to modifications to the \ac{lsst} observing strategy. The observing strategy has shifted from two snaps of 15 second exposures to one 30 second exposures \cite{PSTN-056}. This change relaxes the readout time constraint, and a study of a longer readout sequence was initiated. Three new conditions were tested: longer integration time period of correlated double sampling per pixel, longer readout interval per row, and longer time for parallel charge transfer. After testing at \ac{SLAC} and \ac{UCD}, noise measurements indicated that the modifications to increase the readout interval per row delivered improved read noise performance relative to the \texttt{v30} sequencer, while achieving the targeted 3 second readout time. Additionally, this newly adopted sequencer contained only a minimal deviation from the \texttt{v30} sequencer, preserving parameters that had been previously tuned over years of testing at \ac{SLAC}, \ac{UCD}, and \ac{BNL}. This new sequencer is responsible for the improvement to \ac{e2v} read noise in Table \ref{tab:OnSkyPerformance:Table}. 

As discussed in section \ref{subsubsec:R7_seqOpt}, continuous single pixel read was running while the system was idle as "idle flush", which stabilized bias structure but induced amplifier tearing. An alternative state, idle clear, was implemented for primarily two purposes 1) to keep the readout electronics chain in the same power load (temperature) as possible for realizing a stable gain, and 2) to clear out the accumulated charges continuously so that charge generation in the idle state does not trigger the current limit for the high voltage. The actual implementation is that 100 rows are readout every 1600 ms, which are configurable. The amplitude of thermal fluctuation at the focal plane decreased by $\sim2\%$.


\section{Sensor features and anomalies}\label{sec:CCD_features_anomalies}



During \ac{EOT} and on-sky imaging, we identified a series of sensor defects and anomalies. Here, we describe the identified defects, and the conditions required to produce their effects.

\subsection{Phosphorescence}

After reducing the persistence effect down to a sub-electron level, a second residual charge effect was identified. This residual charge effect affected only a handful of \ac{itl} sensors and was morphologically distinct from the persistence effect. The resulting pattern on affected \ac{itl} sensors is similar to the previously observed 'coffee stain' features, though with opposite polarity.\cite{2023PASP..135k5003E} Consultation with experts at \ac{itl} led to the development of a physical theory driving this second residual charge effect. 

\begin{figure}
    \centering
    \includegraphics[width=\linewidth]{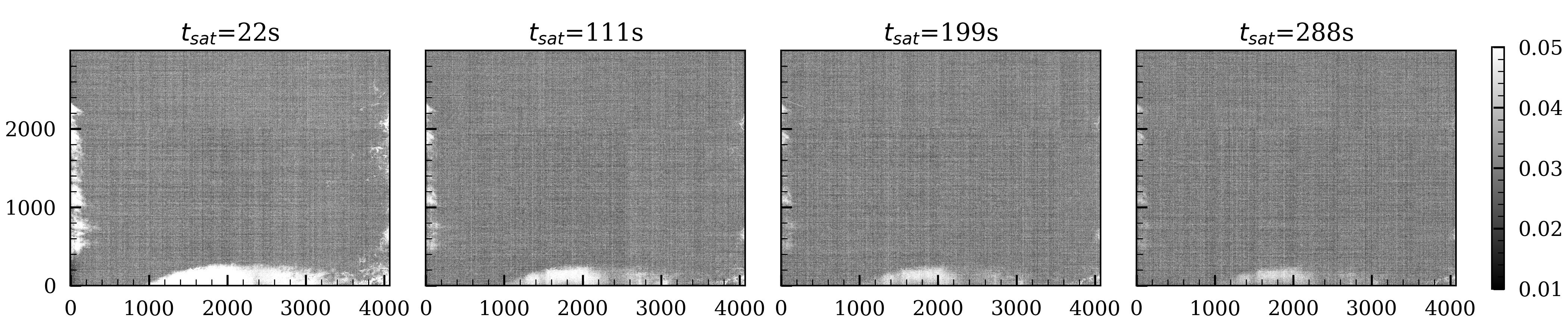}
    \caption{A CCD with a prominent phosphorescent feature, located around the perimeter of the sensor. Fifteen second dark exposures are shown, with the region selected to a $3,000\times4,072$ pixel region localized to the phosphorescence. The time from the initial oversaturated flat ($t_{sat}$) increases moving from left to right. Pixel units are in electrons, and the oversaturated flat was exposed using the 625nm LED of the \ac{CCOB}.}
    \label{fig:phosphorescence}
\end{figure}

After the final silicon acid etch during the CCD fabrication several raised spots on the sensors backside surface were observed. The raised silicon areas could potentially trap the resist which is used during the cleaning process that directly follows the etching step. The resist is wax-based and fluoresces. Due to the longer timescales associated with re-emission, the Rubin team named the feature phosphorescence. An example of the phosphorescence and the decaying effect can be seen in Figure \ref{fig:phosphorescence}.

Several observations were made regarding the phosphorescent signal. The amplitude of phosphorescent response is dependent on the wavelength and amplitude of the illumination, with shorter wavelength light exhibiting a stronger phosphorescence response, as well as higher initial illumination level leading to a higher response. Phosphorescence has two distinct sub-morphologies, diffuse and spot-like. A prominent diffuse phosphorescence is shown in Figure \ref{fig:phosphorescence}. Additionally, the HV back bias affects the phosphorescence signal, though the response is not consistent across different phosphorescent complexes.


The extent of individual phosphorescent complexes in \ac{itl} sensors are static, and we find no evidence for their growth over time. As a result, the image processing treatment involves a defect mask over the largest extent of affected sensor area identified from the \ac{EOT} period.

\subsection{Vampire Pixels}

A unique class of sensor defects was identified as vampire pixels. These defects exhibit a peculiar response pattern: a cental group of pixels with a photo-response exceeding the flat-field mean, surrounded by a concentric distribution of pixels with a photo-response below the same flat-field mean. The over-responsive pixels seem to “suck” signal from the intended receivers, leading to the name vampire pixels. This effect resembles a sort of reverse brighter-fatter effect, excited simply by illumination.

This feature was first identified in \ac{lsstcomcam}, and subsequently found in \ac{lsstcam}. An analysis routine was added to \ac{eo_pipe}\footnote{The eo\_pipe software package (\url{https://github.com/lsst-camera-dh/eo_pipe}) is used to perform a standard set of electro-optical tests on LSSTCam pixel data.} to search for bright defect pixels in combined flats. The resulting distribution confirmed our previous observations, associating the vampire pixel features with \ac{itl} sensors.

\begin{figure}[h!]
    \centering
    \includegraphics[width=0.96\linewidth]{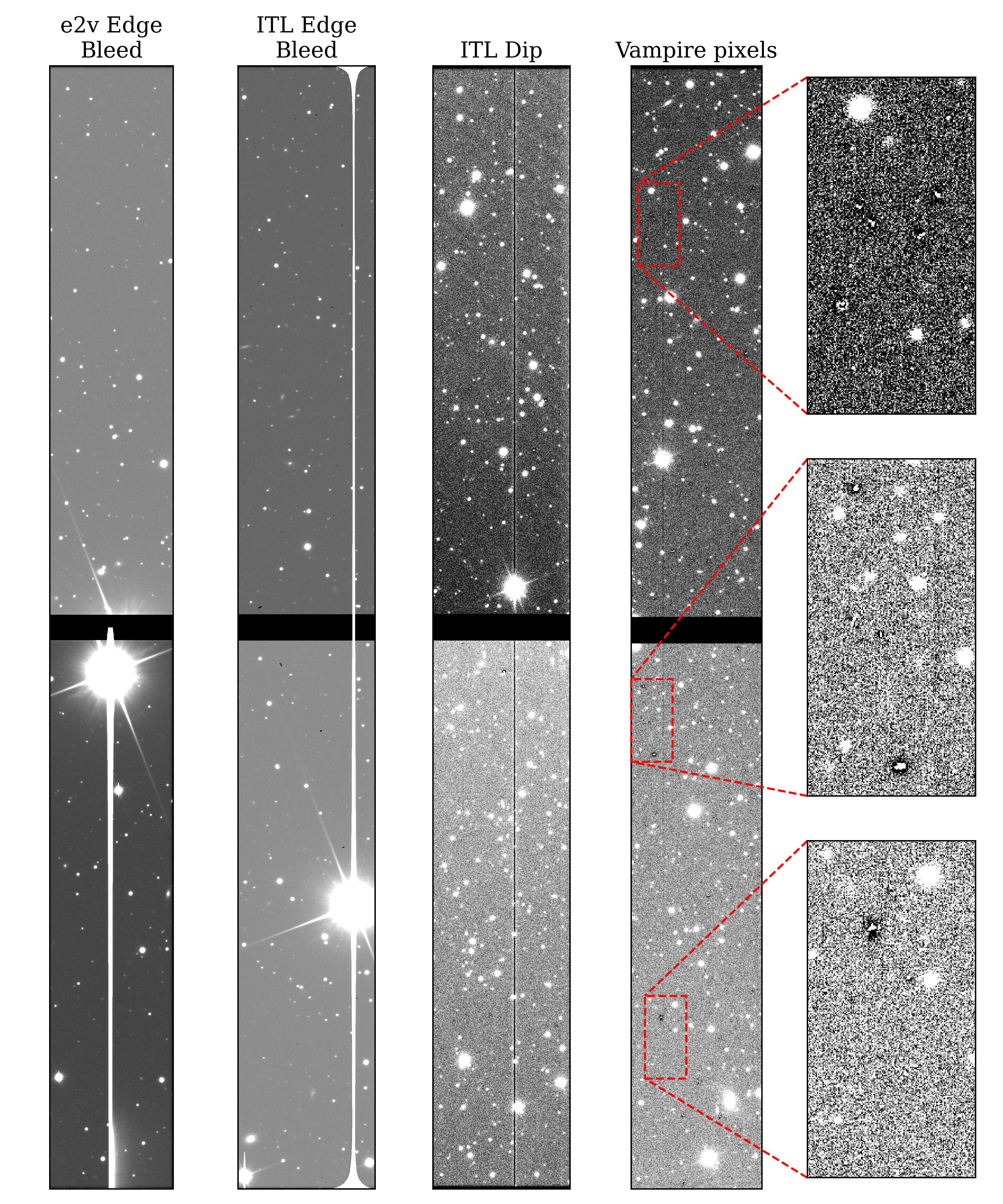}
    \caption{Four of the sensor anomalies observed on-sky with \ac{lsstcam}. All anomalies shown here are from a single on sky exposure from April 11, 2026, with different \ac{ccds} shown for each defect. The inset panels of vampire pixels are 150 pixels wide and 300 pixels long, while the other panels show two adjacent CCD segments and the overscan region.
    }
    \label{fig:defects}
\end{figure}

One observation during our \ac{EOT} campaign was that vampire pixels were a sensor level surface effect, and did not move across the focal plane after reconfiguring the \ac{CCOB} and 4k optical projector. Since some vampire pixels exhibit phosphorescence, we note that several \ac{EOT} and on-sky exposures show a time varying response of the the outer region associated with the lower photo-response, where this region changes in radial size. These vampire pixels have been observed on sky, and several are shown in the right panels of Figure \ref{fig:defects}. We have observed a high number of vampire pixels on two \ac{itl} sensors, which also exhibit high read noise and dark current. Study is ongoing to determine if this correlation is due to the vampire pixels, or other sensor qualities.

The region of the sensor affected by vampire pixels is typically $\sim20$ pixels (200µm) in diameter, and the largest regions being $\sim100$ pixels (1 mm) in diameter. For image processing, we apply a defect mask over the affected sensor area with some buffer to account for any additional pixels that may phosphoresce.

\subsection{\ac{itl} Dips}

\ac{itl} dips were first discovered during commissioning of \ac{lsstcomcam} and later also identified in \ac{lsstcam}. These features are dark columns emanating from bright stars which traverse the entire detector, crossing amplifier boundaries. These dark columns are unique in that the core of the column has a lower signal compared to the adjacent column.

We investigated whether \ac{itl} dips could also be observed in the \ac{ccds} of \ac{lsstcam} during the Run 7 \ac{EOT} period. For this study, we used spots and rectangles projected by the 4K projector onto the focal plane. The spots were approximately 30 pixels (300µm) across, and projected onto every amplifier segment of each detector. We were unable to find any evidence of \ac{itl} dips. Despite the lack of identification during \ac{EOT}, we could not rule out that the feature would not appear on-sky, due to the contrast of the 4k projected rectangles being significantly lower than the contrast from a bright star on the dark sky.

\ac{itl} dips were observed on-sky with \ac{lsstcam}, and we determined that the depth of the dip (central dark column fluctuation) is dependent on the background sky level and the number of significantly saturated neighboring columns. Additionally, we observe that dips with the largest dark column fluctuation relative to sky background exhibit the bright edge columns, but only above a certain detector-dependent threshold (see central sub panel of Figure \ref{fig:defects} for an example). 

The physical model that produces \ac{itl} dips is still not well understood, and several models have been proposed. One such model is dependent on the a change in the density of holes within the channel stops. This effect is the known cause of amplifier tearing effects in \ac{e2v} sensors. To produce the physical effect of an \ac{itl} dip, electrons from saturated stars would overspill across channel stops, lowers the density of holes within the channel stops, and propagating along the channel stops within the time of an exposure (30 seconds) \cite{juramy2026dark}.

While \ac{itl} sensors have different operating conditions than the \ac{e2v} sensors, it remains a plausible physical model, due to the similar effect observed along the channel stops between adjacent \ac{e2v} segments. For on-sky treatment and image processing, the columns are masked as defects. 




\subsection{Edge Bleeds}

Two distinct sensor level effects have been observed in \ac{lsstcam}, with both occurring near the serial registers of \ac{ccds} in response to saturated columns. Despite the localization and anomaly being attributed to an edge bleed, the resulting feature is different for \ac{itl} and \ac{e2v} sensors.

\subsubsection{\ac{itl} sensors}

During commissioning \ac{lsstcomcam}, edge effects near the serial register of \ac{itl} sensors were observed, where columns saturated by bright source would grow in size when in proximity to the serial register. This behavior was also observed in on-sky exposures of \ac{lsstcam}, with the edge bleed region increasing in extent for rows closer to the serial readout register. This behavior is consistent with serial register pixels spilling into the adjacent active area rows during serial clocking. 

During the on-sky sequencer optimization, described in \ref{subsec:onsky_sequencer_idleClear_optimizations}, measurements of \ac{itl} dips were made to look for any correlation with mitigating the \ac{itl} dip, and no clear improvement has been quantified. Despite this, on-sky treatment of \ac{itl} edge dips associated with bright stars are masked as a defect.

\subsubsection{\ac{e2v} sensors}

During the first on-sky exposures of \ac{lsstcam}, different sensor edge effect was identified near the serial register of \ac{e2v} sensors. The defect would manifest as a bright region of pixel $\sim100$ pixels wide, propagating in the opposite direction to the serial readout direction. Notably, these features do not originate at the serial register, and are typically offset by $\sim20-50$ pixels from the serial register, though some \ac{e2v} edge bleeds have been observed adjacent to the serial readout register.

Several hypotheses have been put forward to the source of the \ac{e2v} edge bleed, though no definitive model has been accepted as the physical driver of this defect. The central hypotheses are predicated on saturation in the column contaminating the serial overscan region, which is supported by the observation that the edge bleed is projected opposite of the serial readout register direction. One peculiarity of the defect is that it manifests only on the lower \ac{e2v} segments despite saturated stars frequently being read out on the upper segments. On-sky treatment of \ac{e2v} edge dips is accomplished by masking the edge region.
\section{Conclusion}\label{sec:conclusion}
We summarize the status of the \ac{lsstcam} focal plane as follows:

\begin{itemize}
    \item \ac{lsstcam} was successfully received at Rubin Observatory in May 2024, and underwent a thorough testing period in late 2024 to verify the performance of the \ac{ccds} was satisfactory for Rubin Observatory. During this period, many optimizations were implemented to the readout sequence and thermal management of the focal plane. At the conclusion of this testing period, the \ac{lsstcam} focal plane met its design specification.
    \item \ac{lsstcam} was integrated to the Simonyi Survey Telescope at Rubin Observatory, and first on-sky exposures were achieved on April 15th, 2025. During the integration process, two separate issues related to an electrical short on the focal plane and an anomalous \ac{adc} prevented operation of six science \ac{ccds} from operation for the first on-sky exposures.
    \item The focal plane showed resilience in high-flux testing on-sky, and several other tests were conducted to evaluate the performance of the \ac{ccds}. Additional optimizations were implemented to the readout sequence, aimed at a general reduction in read noise, reduction of long-range noise correlations, and improving CCD safety.
    \item Five of the six \ac{ccds} affected by the electrical short and faulty \ac{adc} were recovered, resulting in 188/189 operational science sensors in \ac{lsstcam}, with no CCD degradation observed during early operations.
    \item Many sensor features and anomalies have been documented and a variety of methods have been employed to mitigate and mask defects in science images. Defect contamination to the active sensor area is deemed to be significantly lower than the system requirement.
\end{itemize}

Rubin Observatory will begin the \ac{lsst} survey in 2026. In preparation for the start of survey, Rubin Observatory is in an early operations period of continued system optimization. The \ac{lsstcam} operational parameters have converged, and we do not anticipate significant operational changes to \ac{lsstcam} ahead of the start of the \ac{lsst} survey. During this period of early operations, no degradation of the \ac{lsstcam} \ac{ccds} has been observed, and the system has demonstrated high standard of performance in the commissioning environment. The operations to date have demonstrated the capability of performing a wide, fast, and deep optical imaging survey of the entire southern sky at the Rubin Observatory, with \ac{lsstcam} serving as the principal instrument to accomplish the \ac{lsst}.



\acknowledgments 
This material is based upon work supported in part by the National Science Foundation through Cooperative Agreements AST-1258333 and AST-2241526 and Cooperative Support Agreements AST-1202910 and AST2211468 managed by the Association of Universities for Research in Astronomy (AURA), and the Department of Energy under Contract No. DE-AC02-76SF00515 with the SLAC National Accelerator Laboratory managed by Stanford University. Additional Rubin Observatory funding comes from private donations, grants to universities, and in-kind support from LSST-DA Institutional Members. This work has been supported by the French National Institute of Nuclear and Particle Physics (IN2P3) through dedicated funding provided by the National Center for Scientific Research (CNRS). This work has been supported by STFC funding for UK participation in LSST, through grant ST/Y00292X/1.

\bibliography{report} 

@TechReport{CTN-001,
    author = "{Plazas Malag\'on}, Andr\'es A. and Digel, Seth W. and Roodman, Aaron and Broughton, Alex and {LSST Camera Team}",
    title = "{LSSTCam and LSSTComCam Focal Plane Layouts}",
    institution = "{NSF-DOE Vera C. Rubin Observatory}",
    year = "2025",
    month = "August",
    handle = "CTN-001",
    type = "{Camera Technical Note}",
    number = "CTN-001",
    doi = "10.71929/rubin/2584019",
    url = "https://ctn-001.lsst.io/"
}

@TechReport{DMTN-276,
    author = "Banovetz, John and Utsumi, Yousuke and Slater, Colin",
    title = "{Effects of Persistence on E2V Sensors and its Impacts on DC2 Data}",
    institution = "{NSF-DOE Vera C. Rubin Observatory}",
    year = "2024",
    month = "September",
    handle = "DMTN-276",
    type = "{Data Management Technical Note}",
    number = "DMTN-276",
    url = "https://dmtn-276.lsst.io/"
}

@TechReport{PSTN-019,
    author = "{Rubin Observatory Science Pipelines Developers}",
    title = "{The LSST Science Pipelines Software: Optical Survey Pipeline Reduction and Analysis Environment}",
    institution = "{NSF-DOE Vera C. Rubin Observatory}",
    year = "2025",
    month = "December",
    handle = "PSTN-019",
    type = "{Project Science Technical Note}",
    number = "PSTN-019",
    doi = "10.71929/rubin/2570545",
    url = "https://pstn-019.lsst.io/"
}

@TechReport{PSTN-056,
    author = "{Rubin's Survey Cadence Optimization Committee} and Bianco, Federica B. and Jones, R. Lynne and Anguita, Timo and Bauer, Franz E. and Edwards, Louise O. V. and Jha, Saurabh W. and Mandelbaum, Rachel and Miller, Adam A. and Olsen, Knut and Slater, Colin T. and Smartt, Stephen J. and Strader, Jay and Street, Rachel A. and Volk, Kathryn and Yoachim, Peter",
    title = "{Survey Cadence Optimization Committee’s Phase 3 Recommendations}",
    institution = "{NSF-DOE Vera C. Rubin Observatory}",
    year = "2025",
    month = "January",
    handle = "PSTN-056",
    type = "{Project Science Technical Note}",
    number = "PSTN-056",
    doi = "10.71929/rubin/2585402",
    url = "https://pstn-056.lsst.io/"
}

@TechReport{SITCOMTN-148,
    author = "{LSST Camera Team} and Antilogus, Pierre and Astier, Pierre and Banovetz, John and Bregeon, Johan and Chiang, James and Digel, Seth W. and Fanning, Kevin and Faris, Yassine and Guillemin, Thibault and Johnson, Anthony S. and Juramy-Gilles, Claire and Lage, Craig S. and Liang, Shuang and MacBride, Sean Patrick and Marshall, Stuart and Neal, Homer and Polin, Daniel and Rasmussen, Andrew P. and Roodman, Aaron and Rykoff, Eli S. and Stalder, Brian and Thayer, John Gregg and Tyson, J. Anthony and Park, Hye Yun and Utsumi, Yousuke and Zhang, Zhuoqi",
    title = "{LSST Camera Electro-Optical Test (Run 7) Results}",
    institution = "{NSF-DOE Vera C. Rubin Observatory}",
    year = "2025",
    month = "August",
    handle = "SITCOMTN-148",
    type = "{Commissioning Technical Note}",
    number = "SITCOMTN-148",
    doi = "10.71929/rubin/2583999",
    url = "https://sitcomtn-148.lsst.io/"
}

@ARTICLE{2019ApJ...873..111I,
    author = "{Ivezi{\'c}}, {\v{Z}}eljko and {Kahn}, Steven M. and {Tyson}, J. Anthony and {Abel}, Bob and {Acosta}, Emily and {Allsman}, Robyn and {Alonso}, David and {AlSayyad}, Yusra and {Anderson}, Scott F. and {Andrew}, John and {Angel}, James Roger P. and {Angeli}, George Z. and {Ansari}, Reza and {Antilogus}, Pierre and {Araujo}, Constanza and {Armstrong}, Robert and {Arndt}, Kirk T. and {Astier}, Pierre and {Aubourg}, {\'E}ric and {Auza}, Nicole and {Axelrod}, Tim S. and {Bard}, Deborah J. and {Barr}, Jeff D. and {Barrau}, Aurelian and {Bartlett}, James G. and {Bauer}, Amanda E. and {Bauman}, Brian J. and {Baumont}, Sylvain and {Bechtol}, Ellen and {Bechtol}, Keith and {Becker}, Andrew C. and {Becla}, Jacek and {Beldica}, Cristina and {Bellavia}, Steve and {Bianco}, Federica B. and {Biswas}, Rahul and {Blanc}, Guillaume and {Blazek}, Jonathan and {Blandford}, Roger D. and {Bloom}, Josh S. and {Bogart}, Joanne and {Bond}, Tim W. and {Booth}, Michael T. and {Borgland}, Anders W. and {Borne}, Kirk and {Bosch}, James F. and {Boutigny}, Dominique and {Brackett}, Craig A. and {Bradshaw}, Andrew and {Brandt}, William Nielsen and {Brown}, Michael E. and {Bullock}, James S. and {Burchat}, Patricia and {Burke}, David L. and {Cagnoli}, Gianpietro and {Calabrese}, Daniel and {Callahan}, Shawn and {Callen}, Alice L. and {Carlin}, Jeffrey L. and {Carlson}, Erin L. and {Chandrasekharan}, Srinivasan and {Charles-Emerson}, Glenaver and {Chesley}, Steve and {Cheu}, Elliott C. and {Chiang}, Hsin-Fang and {Chiang}, James and {Chirino}, Carol and {Chow}, Derek and {Ciardi}, David R. and {Claver}, Charles F. and {Cohen-Tanugi}, Johann and {Cockrum}, Joseph J. and {Coles}, Rebecca and {Connolly}, Andrew J. and {Cook}, Kem H. and {Cooray}, Asantha and {Covey}, Kevin R. and {Cribbs}, Chris and {Cui}, Wei and {Cutri}, Roc and {Daly}, Philip N. and {Daniel}, Scott F. and {Daruich}, Felipe and {Daubard}, Guillaume and {Daues}, Greg and {Dawson}, William and {Delgado}, Francisco and {Dellapenna}, Alfred and {de Peyster}, Robert and {de Val-Borro}, Miguel and {Digel}, Seth W. and {Doherty}, Peter and {Dubois}, Richard and {Dubois-Felsmann}, Gregory P. and {Durech}, Josef and {Economou}, Frossie and {Eifler}, Tim and {Eracleous}, Michael and {Emmons}, Benjamin L. and {Fausti Neto}, Angelo and {Ferguson}, Henry and {Figueroa}, Enrique and {Fisher-Levine}, Merlin and {Focke}, Warren and {Foss}, Michael D. and {Frank}, James and {Freemon}, Michael D. and {Gangler}, Emmanuel and {Gawiser}, Eric and {Geary}, John C. and {Gee}, Perry and {Geha}, Marla and {Gessner}, Charles J. B. and {Gibson}, Robert R. and {Gilmore}, D. Kirk and {Glanzman}, Thomas and {Glick}, William and {Goldina}, Tatiana and {Goldstein}, Daniel A. and {Goodenow}, Iain and {Graham}, Melissa L. and {Gressler}, William J. and {Gris}, Philippe and {Guy}, Leanne P. and {Guyonnet}, Augustin and {Haller}, Gunther and {Harris}, Ron and {Hascall}, Patrick A. and {Haupt}, Justine and {Hernandez}, Fabio and {Herrmann}, Sven and {Hileman}, Edward and {Hoblitt}, Joshua and {Hodgson}, John A. and {Hogan}, Craig and {Howard}, James D. and {Huang}, Dajun and {Huffer}, Michael E. and {Ingraham}, Patrick and {Innes}, Walter R. and {Jacoby}, Suzanne H. and {Jain}, Bhuvnesh and {Jammes}, Fabrice and {Jee}, M. James and {Jenness}, Tim and {Jernigan}, Garrett and {Jevremovi{\'c}}, Darko and {Johns}, Kenneth and {Johnson}, Anthony S. and {Johnson}, Margaret W. G. and {Jones}, R. Lynne and {Juramy-Gilles}, Claire and {Juri{\'c}}, Mario and {Kalirai}, Jason S. and {Kallivayalil}, Nitya J. and {Kalmbach}, Bryce and {Kantor}, Jeffrey P. and {Karst}, Pierre and {Kasliwal}, Mansi M. and {Kelly}, Heather and {Kessler}, Richard and {Kinnison}, Veronica and {Kirkby}, David and {Knox}, Lloyd and {Kotov}, Ivan V. and {Krabbendam}, Victor L. and {Krughoff}, K. Simon and {Kub{\'a}nek}, Petr and {Kuczewski}, John and {Kulkarni}, Shri and {Ku}, John and {Kurita}, Nadine R. and {Lage}, Craig S. and {Lambert}, Ron and {Lange}, Travis and {Langton}, J. Brian and {Le Guillou}, Laurent and {Levine}, Deborah and {Liang}, Ming and {Lim}, Kian-Tat and {Lintott}, Chris J. and {Long}, Kevin E. and {Lopez}, Margaux and {Lotz}, Paul J. and {Lupton}, Robert H. and {Lust}, Nate B. and {MacArthur}, Lauren A. and {Mahabal}, Ashish and {Mandelbaum}, Rachel and {Markiewicz}, Thomas W. and {Marsh}, Darren S. and {Marshall}, Philip J. and {Marshall}, Stuart and {May}, Morgan and {McKercher}, Robert and {McQueen}, Michelle and {Meyers}, Joshua and {Migliore}, Myriam and {Miller}, Michelle and {Mills}, David J.",
    title = "{LSST: From Science Drivers to Reference Design and Anticipated Data Products}",
    journal = "The Astrophysical Journal",
    year = "2019",
    month = "March",
    volume = "873",
    number = "2",
    eid = "111",
    pages = "111",
    doi = "10.3847/1538-4357/ab042c",
    archivePrefix = "arXiv",
    eprint = "0805.2366",
    primaryClass = "astro-ph",
    adsurl = "https://ui.adsabs.harvard.edu/abs/2019ApJ...873..111I"
}

@inproceedings{juramy2026dark,
  author    = {Juramy, Claire and Antilogus, Pierre and Astier, Pierre and Banovetz, John and MacBride, Sean Patrick and Rasmussen, Andrew P. and Utsumi, Yousuke},
  title     = {The ``dark dips'' phenomenon in the {LSST} {Camera} on-sky images},
  year = 2024,
  month = jul,
  series = {X-Ray, Optical, and Infrared Detectors for Astronomy XII},
  year      = {2026},
  note      = {Presented at SPIE Astronomical Telescopes + Instrumentation}
}

@inproceedings{FanningsPaper,
  author    = {Fanning, Kevin},
  title     = {{LSSTCam} Installation and Commissioning at the {Vera C. Rubin Observatory}},
  note         = {SPIE Astronomical Telescopes and Instrumentation 14147-39},
  year      = {2026},
  series    = {Ground-based and Airborne Telescopes XI}
}

@INPROCEEDINGS{2024SPIE13096E..1OL,
       author = {{Lange}, Travis and {Nordby}, Martin and {Pollek}, Hannah and {Osier}, Shawn and {Bowdish}, Boyd and {Hascall}, Diane and {Lopez}, Margaux and {Newbry}, Scott P. and {Lazarte}, Juan Carlos and {Thayer}, Gregg and {Neal}, Homer and {Lee}, Vincent T. and {Silva}, Michael and {Kiehl}, David and {Hau}, Andrew and {Nieland}, Tom and {Linton}, Nico and {Shelley}, David C. and {Qiu}, Yongqiang and {Freytag}, Mark and {Cisneros}, Stephen and {Mendez}, Chris and {Marshall}, Stuart and {Utsumi}, Yousuke and {Johnson}, Tony and {Rasmussen}, Andrew P. and {Roodman}, Aaron and {Tether}, Steven and {Eisner}, Alan and {Turri}, Max and {Onoprienko}, Dmitry and {Saxton}, Owen and {Chiang}, James and {Digel}, Seth W. and {Bradshaw}, Andrew K. and {Reil}, Kevin A. and {Riot}, Vincent and {Wolfe}, Justin E. and {Winters}, Scott and {Bauman}, Brian J. and {Wahl}, William and {O'Connor}, Paul and {Antilogus}, Pierre and {Juramy}, Claire and {Virieux}, Fran{\c{c}}oise and {Boucaud}, Alexandre and {Parisel}, Camille and {Aubourg}, Eric and {Lagorio}, Eric and {Karst}, Pierre and {Marini}, Aurelien and {Laporte}, Didier and {Vezzu}, Francis and {Daubard}, Guillaume and {Breugnon}, Patrick and {Tyson}, Tony and {Snyder}, Adam and {Lage}, Craig and {Bond}, Tim and {Ritz}, Steve},
        title = "{Integrating the LSST camera}",
    booktitle = {Ground-based and Airborne Instrumentation for Astronomy X},
         year = 2024,
       editor = {{Bryant}, Julia J. and {Motohara}, Kentaro and {Vernet}, Jo{\"e}l. R.~D.},
       series = {Society of Photo-Optical Instrumentation Engineers (SPIE) Conference Series},
       volume = {13096},
        month = jul,
          eid = {130961O},
        pages = {130961O},
          doi = {10.1117/12.3019302},
       adsurl = {https://ui.adsabs.harvard.edu/abs/2024SPIE13096E..1OL}
}

@INPROCEEDINGS{2024SPIE13096E..1SR,
       author = {{Roodman}, A. and {Rasmussen}, A. and {Bradshaw}, A. and {Charles}, E. and {Chiang}, J. and {Digel}, S.~W. and {Dubois}, R. and {Johnson}, A.~S. and {Kahn}, S. and {Liang}, S. and {Marshall}, S. and {Neal}, H. and {Plazas}, A.~A. and {Reil}, K. and {Rykoff}, E. and {Schindler}, R. and {Schutt}, T. and {Utsumi}, Y. and {Bogart}, T. and {Bond}, T. and {Bowdish}, B. and {Cisneros}, S. and {Eisner}, A. and {Freytag}, M. and {Hascall}, D. and {Lange}, T. and {Lazarte}, J.~C. and {Lopez}, M. and {Mendez}, C. and {Newbry}, S. and {Nordby}, M. and {Onoprienko}, D. and {Osier}, S. and {Pollek}, H. and {Qiu}, B. and {Saxton}, O. and {Tether}, S. and {Thayer}, G. and {Turri}, M. and {Banovetz}, J. and {O'Connor}, P. and {Riot}, V. and {Wolfe}, J. and {Lage}, C. and {Polin}, D. and {Snyder}, A. and {Tyson}, A. and {Nichols}, R. and {Ritz}, S. and {Shestakov}, A. and {Wood}, D. and {Broughton}, A. and {Park}, H. and {Esteves}, J. and {Barrau}, A. and {Bregeon}, J. and {Combet}, C. and {Dargaud}, G. and {Lagorio}, E. and {Migliore}, M. and {Vezzu}, F. and {Antilogus}, P. and {Astier}, P. and {Daubard}, G. and {Juramy}, C. and {Laporte}, D. and {Guillemin}, T. and {Aubourg}, E. and {Boucaud}, A. and {Parisel}, C. and {Virieux}, F. and {Breugnon}, P. and {Karst}, P. and {Marini}, A. and {Fisher-Levine}, M. and {Waters}, C.},
        title = "{LSST camera verification testing and characterization}",
    booktitle = {Ground-based and Airborne Instrumentation for Astronomy X},
         year = 2024,
       editor = {{Bryant}, Julia J. and {Motohara}, Kentaro and {Vernet}, Jo{\"e}l. R.~D.},
       series = {Society of Photo-Optical Instrumentation Engineers (SPIE) Conference Series},
       volume = {13096},
        month = jul,
          eid = {130961S},
        pages = {130961S},
          doi = {10.1117/12.3019698},
       adsurl = {https://ui.adsabs.harvard.edu/abs/2024SPIE13096E..1SR}
}

@INPROCEEDINGS{2014SPIE.9154E..15T,
       author = {{Tyson}, J.~A. and {Sasian}, J. and {Gilmore}, K. and {Bradshaw}, A. and {Claver}, C. and {Klint}, M. and {Muller}, G. and {Poczulp}, G. and {Resseguie}, E.},
        title = "{LSST optical beam simulator}",
    booktitle = {High Energy, Optical, and Infrared Detectors for Astronomy VI},
         year = 2014,
       editor = {{Holland}, Andrew D. and {Beletic}, James},
       series = {Society of Photo-Optical Instrumentation Engineers (SPIE) Conference Series},
       volume = {9154},
        month = jul,
          eid = {915415},
        pages = {915415},
          doi = {10.1117/12.2055604},
archivePrefix = {arXiv},
       eprint = {1411.5667},
 primaryClass = {astro-ph.IM},
       adsurl = {https://ui.adsabs.harvard.edu/abs/2014SPIE.9154E..15T}
}

@INPROCEEDINGS{2014SPIE.9147E..74R,
       author = {{Riot}, Vincent J. and {Arndt}, Kirk and {Claver}, Chuck and {Doherty}, Peter E. and {Gilmore}, D.~K. and {Hascall}, Patrick A. and {Herrmann}, Sven and {Kotov}, Ivan and {O'Connor}, Paul and {Sebag}, Jacques and {Stubbs}, Christopher W. and {Warner}, Michael},
        title = "{The guider and wavefront curvature sensor subsystem for the Large Synoptic Survey Telescope}",
    booktitle = {Ground-based and Airborne Instrumentation for Astronomy V},
         year = 2014,
       editor = {{Ramsay}, Suzanne K. and {McLean}, Ian S. and {Takami}, Hideki},
       series = {Society of Photo-Optical Instrumentation Engineers (SPIE) Conference Series},
       volume = {9147},
        month = aug,
          eid = {914774},
        pages = {914774},
          doi = {10.1117/12.2056605},
       adsurl = {https://ui.adsabs.harvard.edu/abs/2014SPIE.9147E..74R}
}

@INPROCEEDINGS{2018SPIE10702E..2CL,
       author = {{Lopez}, Margaux and {Marshall}, Stuart and {Bond}, Tim and {Haupt}, Justine and {Johnson}, Tony and {Neal}, Homer and {O'Connor}, Paul and {Rasmussen}, Andrew and {Roodman}, Aaron and {Takacs}, Peter and {Utsumi}, Yousuke},
        title = "{Acceptance testing for LSST camera raft tower modules}",
    booktitle = {Ground-based and Airborne Instrumentation for Astronomy VII},
         year = 2018,
       editor = {{Evans}, Christopher J. and {Simard}, Luc and {Takami}, Hideki},
       series = {Society of Photo-Optical Instrumentation Engineers (SPIE) Conference Series},
       volume = {10702},
        month = jul,
          eid = {107022C},
        pages = {107022C},
          doi = {10.1117/12.2312200},
       adsurl = {https://ui.adsabs.harvard.edu/abs/2018SPIE10702E..2CL}
}

@ARTICLE{2020arXiv200103223S,
       author = {{Snyder}, Adam and {Roodman}, Aaron},
        title = "{Investigation of Deferred Charge Effects in LSST ITL Sensors}",
      journal = {arXiv e-prints},
         year = 2020,
        month = jan,
          eid = {arXiv:2001.03223},
        pages = {arXiv:2001.03223},
          doi = {10.48550/arXiv.2001.03223},
archivePrefix = {arXiv},
       eprint = {2001.03223},
 primaryClass = {astro-ph.IM},
       adsurl = {https://ui.adsabs.harvard.edu/abs/2020arXiv200103223S}
}

@ARTICLE{2024PASP..136d5003B,
       author = {{Broughton}, Alex and {Utsumi}, Yousuke and {Plazas Malag{\'o}n}, Andr{\'e}s A. and {Waters}, Christopher and {Lage}, Craig and {Snyder}, Adam and {Rasmussen}, Andrew and {Marshall}, Stuart and {Chiang}, Jim and {Murgia}, Simona and {Roodman}, Aaron},
        title = "{Mitigation of the Brighter-fatter Effect in the LSST Camera}",
      journal = {Publications of the Astronomical Society of the Pacific},
         year = 2024,
        month = apr,
       volume = {136},
       number = {4},
          eid = {045003},
        pages = {045003},
          doi = {10.1088/1538-3873/ad3aa2},
archivePrefix = {arXiv},
       eprint = {2312.03115},
 primaryClass = {astro-ph.IM},
       adsurl = {https://ui.adsabs.harvard.edu/abs/2024PASP..136d5003B}
}

@article{astier2019shape,
  title={The shape of the photon transfer curve of CCD sensors},
  author={Astier, Pierre and Antilogus, Pierre and Juramy, Claire and Le Breton, R{\'e}my and Le Guillou, Laurent and Sepulveda, Eduardo and Dark Energy Science Collaboration},
  journal={Astronomy \& Astrophysics},
  volume={629},
  pages={A36},
  year={2019},
  publisher={EDP Sciences}
}

@ARTICLE{2020arXiv200209439J,
       author = {{Juramy}, Claire and {Antilogus}, Pierre and {Le Guillou}, Laurent and {Sepulveda}, Eduardo},
        title = "{Tearing and related field distortions in deep-depletion CCDs}",
      journal = {arXiv e-prints},
         year = 2020,
        month = feb,
          eid = {arXiv:2002.09439},
        pages = {arXiv:2002.09439},
          doi = {10.48550/arXiv.2002.09439},
archivePrefix = {arXiv},
       eprint = {2002.09439},
 primaryClass = {astro-ph.IM},
       adsurl = {https://ui.adsabs.harvard.edu/abs/2020arXiv200209439J}
}

@inproceedings{DOHERTY,
author = {Peter E. Doherty and Pierre Antilogus and Pierre Astier and James Chiang and D. Kirk Gilmore and Augustin Guyonnet and Dajun Huang and Heather Kelly and Ivan Kotov and Petr Kubanek and Andrei Nomerotski and Paul O’Connor and Andrew Rasmussen and Vincent J. Riot and Christopher W. Stubbs and Peter Takacs and J. Anthony Tyson and Kurt Vetter},
title = {{Electro-optical testing of fully depleted CCD image sensors for the Large Synoptic Survey Telescope camera}},
volume = {9154},
booktitle = {High Energy, Optical, and Infrared Detectors for Astronomy VI},
editor = {Andrew D. Holland and James Beletic},
organization = {International Society for Optics and Photonics},
publisher = {SPIE},
pages = {915418},
year = {2014},
doi = {10.1117/12.2056733},
URL = {https://doi.org/10.1117/12.2056733}
}

@INPROCEEDINGS{2024SPIE13103E..0WU,
       author = {{Utsumi}, Yousuke and {Antilogus}, Pierre and {Astier}, Pierre and {Banovetz}, John and {Bradshaw}, Andrew K. and {Bregeon}, Johan and {Broughton}, Alex and {Chiang}, Jim and {Combet}, Celine and {Dargaud}, Guillaume and {Digel}, Seth W. and {Esteves}, Johnny and {Guillemin}, Thibault and {Johnson}, Tony and {Juramy}, Claire and {Lage}, Craig and {Liang}, Shuang and {Marshall}, Stuart and {Migliore}, Myriam and {Neal}, Homer and {Nichols}, Renee and {Polin}, Daniel and {Rasmussen}, Andrew P. and {Ritz}, Steve and {Rykoff}, Eli and {Roodman}, Aaron and {Schutt}, Theo and {Sheshtakov}, Adrian and {Snyder}, Adam and {Thayer}, Gregg and {Turri}, Max and {Tyson}, Tony and {Wood}, Duncan},
        title = "{LSST Camera focal plane optimization}",
    booktitle = {X-Ray, Optical, and Infrared Detectors for Astronomy XI},
         year = 2024,
       editor = {{Holland}, Andrew D. and {Minoglou}, Kyriaki},
       series = {Society of Photo-Optical Instrumentation Engineers (SPIE) Conference Series},
       volume = {13103},
        month = aug,
          eid = {131030W},
        pages = {131030W},
          doi = {10.1117/12.3019117},
       adsurl = {https://ui.adsabs.harvard.edu/abs/2024SPIE13103E..0WU}
}

@ARTICLE{2025JInst..20P7031P,
       author = {{Polin}, Daniel and {Snyder}, Adam and {Lage}, Craig and {Anthony Tyson}, J.},
        title = "{Characterization of residual charge images in LSST camera e2v CCDs}",
      journal = {Journal of Instrumentation},
         year = 2025,
        month = jul,
       volume = {20},
       number = {7},
          eid = {P07031},
        pages = {P07031},
          doi = {10.1088/1748-0221/20/07/P07031},
archivePrefix = {arXiv},
       eprint = {2502.05418},
 primaryClass = {astro-ph.IM},
       adsurl = {https://ui.adsabs.harvard.edu/abs/2025JInst..20P7031P}
}

@ARTICLE{2025JATIS..11a1205B,
       author = {{Banovetz}, John and {Utsumi}, Yousuke and {Meyers}, Joshua and {Beleznay}, Maya and {Rasmussen}, Andrew and {Roodman}, Aaron},
        title = "{``Weather'' in the LSST camera: investigating patterns in differenced flat images}",
      journal = {Journal of Astronomical Telescopes, Instruments, and Systems},
         year = 2025,
        month = jan,
       volume = {11},
          eid = {011205},
        pages = {011205},
          doi = {10.1117/1.JATIS.11.1.011205},
archivePrefix = {arXiv},
       eprint = {2411.13386},
 primaryClass = {astro-ph.IM},
       adsurl = {https://ui.adsabs.harvard.edu/abs/2025JATIS..11a1205B}
}

@ARTICLE{2023PASP..135k5003E,
       author = {{Esteves}, Johnny H. and {Utsumi}, Yousuke and {Snyder}, Adam and {Schutt}, Theo and {Broughton}, Alex and {Trbalic}, Bahrudin and {Mau}, Sidney and {Rasmussen}, Andrew and {Plazas Malag{\'o}n}, Andr{\'e}s A. and {Bradshaw}, Andrew and {Marshall}, Stuart and {Digel}, Seth and {Chiang}, James and {Rykoff}, Eli and {Waters}, Chris and {Soares-Santos}, Marcelle and {Roodman}, Aaron},
        title = "{Photometry, Centroid and Point-spread Function Measurements in the LSST Camera Focal Plane Using Artificial Stars}",
      journal = {Publications of the Astronomical Society of the Pacific},
         year = 2023,
        month = nov,
       volume = {135},
       number = {1053},
          eid = {115003},
        pages = {115003},
          doi = {10.1088/1538-3873/ad0a73},
archivePrefix = {arXiv},
       eprint = {2308.00919},
 primaryClass = {astro-ph.IM},
       adsurl = {https://ui.adsabs.harvard.edu/abs/2023PASP..135k5003E}
}
\bibliographystyle{spiebib} 

\end{document}